\documentclass[onecolumn,showpacs,preprintnumbers,amsmath,amssymb,5pt]{revtex4-2}
\usepackage{graphicx}
\usepackage{epsfig,graphics}
\usepackage{dcolumn}
\usepackage{bm}
\usepackage{multirow}
\usepackage{color}
\usepackage{mathrsfs,amsmath}
\usepackage{mathrsfs}
\usepackage{mhchem}
\usepackage{mathtools}

\begin{document}
\newtheorem{thm}{Theorem}[section]
\newtheorem{lemma}[thm]{Lemma}
\newtheorem{prop}[thm]{Proposition}
\newtheorem{rem}[thm]{Remark}
\newtheorem{cor}[thm]{Corollary}

\title{Thermodynamics for Reduced Models of Chemical Reactions by PEA and QSSA}
 \author{Liangrong Peng}
 \email{peng@mju.edu.cn} 
 \affiliation{College of Mathematics and Data Science, Minjiang University, Fuzhou, 350108, P.R. China.}
 \author{Liu Hong}%
 \email{hongliu@sysu.edu.cn}
 \affiliation{%
 	School of Mathematics, Sun Yat-Sen University, Guangzhou, 510275, P.R. China.
 }%

\begin{abstract}
Partial equilibrium approximation (PEA) and quasi-steady-state approximation (QSSA) are two classical methods for reducing complex macroscopic chemical reactions into simple computable ones. Previous studies mainly focus on the accuracy of solutions before and after applying model reduction. While, in this paper we start from a thermodynamic view, and try to establish a quantitative connection on the essential thermodynamic quantities, like entropy production rate, free energy dissipation rate and entropy flow rate, between the original reversible chemical mass-action equations and the reduced models by either PEA or QSSA. Our results reveal that the PEA and QSSA do not necessarily preserve the nice thermodynamic structure of the original full model during the reduction procedure (e.g. the loss of non-negativity of free energy dissipation rate), especially when adopting the algebraic relations in replace of differential equations. These results are  further validated though the application to Michaelis-Menten reactions analytically and numerically as a prototype. We expect our study would motivate a re-examination on the effectiveness of various model reduction or approximation methods from a new perspective of non-equilibrium thermodynamics.\\

\end{abstract}

\maketitle
\date{today}

\section{Introduction}
Physical systems usually contain multiple time and length scales. Typical examples include galaxy formation, complex fluids, chemical reaction networks (CRNs), and so on. For each time scale and length scale, different descriptions are generally adopted correspondingly in order to grasp key features of the system and to avoid extra complications in modeling. As an example, Navier-Stokes equations have been proven quite satisfactory for simple gas flows in macroscopic scales, while Boltzmann equations are more preferred in meso- and nano-scales. Generally speaking, the microscopic description contains more comprehensive details and at the same time involves a greater computational cost than the macroscopic description. 

Although descriptions in various scales may vary significantly, it is expected that we can go from one description to another through the procedure of coarse-graining in principle, since we are dealing with the same phenomenon and the same physical system. During this procedure, a common thermodynamic structure, including both the conservation of various time-invariant quantities, like density, momentum and energy, and the dissipative structure, expressed usually through the non-decreasing of entropy or non-negativeness of entropy production rate, could be identified at each level of descriptions as a consequence of laws of thermodynamics. Therefore we pursue some key questions in non-equilibrium thermodynamics, i.e., what is the influence of the coarse-graining on the prediction of the dynamical and thermodynamic properties of a system; does there exist a common thermodynamic structure, and if the answer is positive, how it could be well preserved among different levels of descriptions. 

Chemical reactions are ubiquitous in various disciplines, such as biology, chemical engineering, and material science, due to their involvement in matter synthesis and decomposition, energy storage and release, as well as biofunction activation and deactivation 
\cite{segel1989quasi,wachtel2018thermodynamically,voorsluijs2020thermodynamic}.
CRNs are typical systems containing different time and length scales because of the intrinsic complexity and non-linearity of reactions. The descriptions of CRNs can be deterministic or stochastic depending on the amplitude of fluctuations of particle numbers. The stochastic dynamics are usually modeled by chemical master equations, Langevin equations, or Fokker-Planck equations in mesoscopic scales, while deterministic dynamics are modeled by mass-action equations in the macroscopic scale. 

There arises a growing body of results \cite{pigolotti2008coarse, puglisi2010entropy,esposito2012stochastic,kawaguchi2013fluctuation,ge2016mesoscopic,ge2017mathematical,bo2017multiple,rao2018conservation,yong2012, Peng2018Generalized,busiello2019entropy,busiello2019entropynpj,voorsluijs2020thermodynamic, yu2021inverse, yu2022state} in the thermodynamic quantities for (chemical) master equations, the (corresponding coarse-grained) Fokker-Planck equations, and the mass-action equations. 
As to master equations, Puglisi et al. \cite{puglisi2010entropy} demonstrated that the disappearance of loops during coarse-graining results in a reduced entropy production rate. Esposito et al. \cite{esposito2012stochastic,rao2018conservation} provided a framework on stochastic thermodynamics under coarse-graining. It was derived that the entropy production may increase upon coarse-graining when even and odd variables coexist under time reversal \cite{kawaguchi2013fluctuation}. Bo and Celani \cite{bo2017multiple} showed that by employing asymptotic techniques, it is unattainable to provide an accurate depiction of thermodynamics solely based on the slow variables for multiple-scale stochastic processes. 
Despite that master equations can be effectively modeled by Fokker–Planck equations in a coarse-grained description, it was revealed that the entropy production contains valuable information regarding microscopic currents which are not accounted for by the Fokker–Planck equations \cite{busiello2019entropy,busiello2019entropynpj}. 
Consequently, a new method for coarse-graining a microscopic system that has the potential to encompass all the information overlooked in the conventional Kramers-Moyal expansion was introduced \cite{busiello2019entropynpj}. 
As the studies \cite{ge2016mesoscopic,ge2017mathematical} claimed, the same thermodynamic structure, which is purely determined by the physical system under study, will be preserved in both macroscopic and mesoscopic descriptions. To be exact, under the conditions of complex balance, there is a well-defined one-to-one correspondence between mesoscopic and macroscopic descriptions of the same chemical reaction system in the limit of large volume, $V\rightarrow \infty$, not only for the free energy but also for the time evolution of it. Furthermore, it was shown \cite{yong2012, Peng2018Generalized} that a mathematically elegant dissipative structure, expressed through the famous Onsager's reciprocal relations between non-equilibrium forces and fluxes are well preserved during the procedure of coarse-graining, under the detailed-balance condition. 
When further constrained to CRNs, Yu et al. \cite{yu2021inverse} observed that the energy dissipation rate exhibits an inverse power-law relationship with the number of microscopic states within a coarse-grained state. Later, a framework called State-Space Renormalization Group \cite{yu2022state}, was proposed to uncover the evolution of the correlation function between net probability fluxes in CRNs subjected to coarse-graining. For  mass-action equations, Voorsluijs et al. \cite{voorsluijs2020thermodynamic} studied the thermodynamic validity criterion for the irreversible Michaelis-Menten (MM) reactions at the steady state, which guaranteed the irreversible approximation to be quantitatively reliable for a certain range of parameters.

To be concrete, in this paper, we are going to examine a chemical reaction system constituted by massive particles, which makes it a macroscopic system in nature. The reactions are assumed to proceed in a closed reaction vessel containing rapidly stirred chemical solutions at constant temperature $\mathcal{T}$ and constant pressure $\mathcal{P}$. Without loss of generality, we suppose there are $N$ species $\{S_1,S_2,\cdots,S_N\}$ participating in the following reversible reactions,
\begin{equation}
\label{r}
\nu _{i1} ^ +  S_{{1}} {\text{ + }}\nu _{i{{2}}} ^ +  S_{{2}} {\text{ + }} \cdots {\text{ + }}\nu _{iN} ^ +  S_{{N}}
\xrightleftharpoons[{\kappa_i^-}]{{\kappa_i^+}}
\nu _{i1} ^ -  S_{{1}} {\text{ + }}\nu _{i{\text{2}}} ^ -  S_{{2}} {\text{ + }} \cdots {\text{ + }}\nu _{iN} ^ -  S_{{N}},
\end{equation}
where $\nu _{ik} ^ +\geq 0$ and $\nu _{ik} ^ -\geq 0$ are stoichiometric numbers, $\nu _{ik}=\nu _{ik} ^ +- \nu _{ik} ^ -$,
$\kappa _i^+\geq 0$ and $\kappa _i ^-\geq 0$ are the forward and backward rate constants, of the $i$th reaction ($i=1,2,\cdots,M$). The reversibility of  reactions indicates that $\kappa _i^+ >0$ if and only if $\kappa _i^- >0$. 

It is well known that, if the law of mass-action is assumed, the time evolution of the current system is governed by chemical mass-action equations in the macroscopic scale. On the other hand, one can also make a more detailed description of the system in a mesoscopic way. 
In that case, one can use the molecular number of $N$ species $\{\tilde n_1, \tilde n_2, \cdots, \tilde n_N\}$ in a given volume $V$ as random variables to describe the states of the system. The probability of the system in the state $\vec{n}=(n_1, n_2,\cdots, n_N)$ at time $t$ is denoted as $P_t(\vec{n})=P(n_1, n_2, \cdots, n_N; t)$. Then the reaction system in \eqref{r} is characterized by the chemical master equations.

Different from previous studies, our current work concerns about the applicability of methods of model reduction, such as partial equilibrium approximation (PEA) and quasi-steady-state approximation (QSSA), from the aspect of thermodynamics. For general reversible chemical reactions characterized by macroscopic mass-action laws, the relation between the thermodynamic structure before and after model reduction by using either PEA or QSSA is established. In particular, formulas for the entropy flow rate, entropy production rate and free energy dissipation rate of the reduced models of chemical reactions are derived for both PEA and QSSA, respectively. The application of our general theory to the MM reactions further highlights the necessity and complexity of our current study.

The rest paper is organized as follows. In Sec. II, the mass-action equations for macroscopic chemical reactions are introduced, along with the PEA and QSSA methods. Section III contains our main results, including the non-equilibrium thermodynamics for the full chemical reactions, and the thermodynamics for the reduced models by PEA and QSSA methods separately. The MM reactions are studied in detail as a concrete example in Sec. IV, which highlights the complexity of model reduction thermodynamics in the presence of conservative quantities. The last section is the discussion and conclusion.

\section{General theory for closed chemical reactions}

\subsection{Chemical mass-action equations in macroscopic scale}
When going to the macroscopic scale, we consider the case that both the system volume and the molecular number are large enough while their relative ratio is kept constant. Mathematically, we assume the system satisfies the following conditions
$\lim_{V,N \rightarrow\infty}{n_k}/{V}=constant$ for $k=1,2, \cdots, N$.  
Here in the macroscopic scale, we can define the molecular concentration of each species with respect to mesoscopic probability function $P_t(\vec{n})$ as $  c_k(t)=\lim\limits_{V\rightarrow \infty} \sum\limits_{n \in \Omega} V^{-1}n_kP_t(\vec{n})$, where $\Omega$ denotes the state space,   
whose time evolution is characterized through the following mass-action equations \cite{van1983},
\begin{equation}
\label{massactioneq}
\frac{{d}}{{dt}}c_k(t) = \sum\limits_{i = 1}^M {(\nu _{ik} ^ -   - \nu _{ik} ^ +  )} \left( {R_i ^ +  (\vec{c}) - R_i ^ -  (\vec{c})} \right),\quad c_k(0)=c_k^0, \quad k = 1,2,...,N,
\end{equation}
with the respective forward and backward macroscopic reaction rate functions, of the polynomial form, being defined as 
\begin{equation}
\label{massactionlaw}
R_i^+(\vec{c}) = \kappa _i ^ +  \prod\limits_{j = 1}^N {c_j ^{\nu _{ij} ^ +  } } ,\;\;\;R_i^-(\vec{c}) = \kappa _i ^ -  \prod\limits_{j = 1}^N {c_j ^{\nu _{ij} ^ -  } }.
\end{equation}

According to the convergence theorem proven by Kurtz in 1970s \cite{Kurtz1972}, for any finite time, the stochastic processes $\{V^{-1}\tilde n_k\}$ will converge in probability to solutions of the mass-action equations $\{c_k\}$ as $V \rightarrow \infty$, if initial conditions satisfy $\lim_{V \rightarrow \infty} V^{-1} \tilde n_k(0) = c_k^0$ for each $k~(k=1,2, \cdots, N)$. Kurtz's theorem pointed out the intrinsic correspondence between chemical master equations and chemical mass-action equations in different scales. When the system volume and the molecular number are large enough while their relative ratio is kept constant, the mesoscopic and the macroscopic models predict the same result for the same physical system as expected.

The \emph{principle of detailed balance} guarantees that a chemical reaction network, in the thermodynamic equilibrium, must have zero net flux in each pair of forward and backward reactions, $R_i^+(\vec{c})-R_i^-(\vec{c})=0, ~(i = 1,2,...,M)$ \cite{qian2021stochastic}. Mathematically, the mass-action system in Eq. \eqref{massactioneq} is with detailed balance if and only if there exists a positive static state (or equilibrium state) $\vec{c^e}=(c^e_1 , c^e_2, \cdots , c^e_N)^T>0$ such that 
\begin{equation}
\label{macrocomplex}
\kappa _i ^ +  \prod\limits_{j = 1}^N {(c_j ^{e} )^{\nu _{ij} ^ +  } } = \kappa _i ^ -  \prod\limits_{j = 1}^N {(c_j ^{e} )^{\nu _{ij} ^ -  } },\quad i = 1,2,...,M.
\end{equation}


\subsection{Model reduction for chemical reactions}
With the increase of either reactions or reactants, the chemical reaction dynamics becomes complicated in a rapid way. 
Classical methods of model reduction for chemical reactions include PEA, QSSA, maximum entropy principle, and etc. There have been plenty of discussions on the sufficient conditions for the rigorous convergence of PEA and QSSA solutions \cite{kokotovic1999singular,yong2012}. In this section, We will present the basic ideas of PEA and QSSA first, before applying them to derive the simplified model for the chemical mass-action equations. 

\subsubsection{Partial equilibrium approximation}
As to all $M$ reactions, PEA assumes that some reversible ones (saying, the first $W$ reactions) take less time to reach (partial) equilibrium than the others. These reversible reactions are called fast reactions. After reaching equilibrium, the fast reactions make no contribution to the evolution of concentration for each reactant \cite{gorban2005invariant}. By introducing a small parameter $0<\epsilon\ll 1$ to character the fastness of these reactions, which equals to the ratio of relaxation times of the fast and slow reactions, we can recast the ordinary differential equations (ODEs) in \eqref{massactioneq} into  
\begin{equation}
\label{PEA0}
\frac{{d}}{{dt}}c_k(t) = 
-\frac{1}{\epsilon}\sum\limits_{i = 1}^W {\nu _{ik}} \left({\widehat{R_i^+}(\vec{c}) - \widehat{R_i^-}(\vec{c})} \right) -
\sum\limits_{i = W + 1}^M {\nu _{ik}} \left( {R_i ^ +  (\vec{c}) - R_i ^ -  (\vec{c})} \right), \quad c_k(0)=c_k^0, \quad k = 1,2,...,N, 
\end{equation}
where $\widehat{R_i^{\pm}}(\vec{c})$ is of the same order for the fast reactions $(i=1,2,\cdots,W)$ as those for the slow reactions $(i=W+1,W+2,\cdots,M)$ by re-scaling the fast ones as $\widehat{R_i^{\pm}}(\vec{c}) \equiv {\epsilon}R_i^{\pm}(\vec{c})$ $(i=1,2,\cdots,W)$. 

Using PEA, or equivalently, in the limit of $\epsilon\rightarrow0$, Eq. \eqref{PEA0} degenerates into $W$ algebraic equations and $(N-V)$ ODEs as 
\begin{subequations}
\label{PEA1and2}
\begin{align}
\label{PEA1}
 &R_i ^ +  (\vec{c}) - R_i ^ -  (\vec{c})=0,\;\;\;i = 1,2,...,W, \\ 
\label{PEA2}
 &\frac{{d}}{{dt}}c_k(t) =-\sum\limits_{i = W + 1}^M \nu _{ik} \left( {R_i ^ +  (\vec{c}) - R_i ^ -  (\vec{c})} \right),\quad c_k(0)=c_k^0, \quad k = V+1,V+2,...,N. 
  \end{align}
\end{subequations}
Here we assume the concentrations of the first $V$ ($V\leq W$) species could be explicitly solved from Eq. \eqref{PEA1} as functions of the remaining $(N-V)$ variables.  To be concrete, we take logarithmic transformation on both sides of Eq. \eqref{PEA1}, which leads to
\begin{equation*}
\underbrace{ \sum_{j=1}^V \nu_{ij}\ln c_j }_{\text{the i-th element of}~A\vec{x}}
=\underbrace{ \ln \left(\frac{\kappa _i ^ -}{\kappa _i ^ +} \right)-\sum_{j=V+1}^N \nu_{ij}\ln c_j }_{\text{the i-th element of}~\vec{b}_1}, \quad i = 1,2,...,W.
\end{equation*}
Rewrite it into a matrix form $A\vec{x}=\vec{b}_1$, where the rank of matrix $A=[(\nu_{ij})]_{W\times V}$ is assumed to be $V$, $\vec{x}=(\ln c_1,\cdots, \ln c_V)^T$ and $\vec{b}_1=(\ln (\kappa _1 ^ -/\kappa _1 ^ +)-\sum_{j=V+1}^N \nu_{1j}\ln c_j,\cdots, \ln (\kappa _W ^ -/\kappa _W ^ +)-\sum_{j=V+1}^N \nu_{Wj}\ln c_j)^T$ are both vectors of size $V$ and $W$ respectively. Since $A$ is non-degenerate, the vector $\vec{x}$ could be explicitly solved,  $\vec{x}=(A^TA)^{-1}A^T\vec{b}_1$, which means 
\begin{equation}
(c_1,\cdots,c_V)^T=\exp[(A^TA)^{-1}A^T\vec{b}_1].
\end{equation}

Besides solving the algebraic equations, an alternative way is to remove the contribution of fast reactions directly from the ODEs. Correspondingly, the reduced system contains only slow reactions and becomes easily solvable, i.e.
\begin{equation}
\label{PEA4}
\begin{array}{l}
\frac{{d}}{{dt}}c_k(t) =  -
\sum\limits_{i = W + 1}^M {\nu _{ik}} \left( {R_i ^ +  (\vec{c}) - R_i ^ -  (\vec{c})} \right), \quad c_k(0)=c_k^0, \quad k = 1,2,...,N.
 \end{array}
\end{equation}
For clarity, we call Eq. \eqref{PEA1and2} as PEA-1 and Eq. \eqref{PEA4} as PEA-2. We mention that in PEA the initial layer phenomenon emerges and may cause an inconsistency in the initial values between the reduced model and the original full one. Mathematically, since the small parameter $\epsilon$ appears in the highest order of Eq. \eqref{PEA0}, solving the approximation solutions becomes a singular perturbation problem \cite{huang2015partial}. Similar arguments are applicable to QSSA too. 

\subsubsection{Quasi-steady-state approximation}
\label{Quasi steady state approximation}
In contrast to rapid balance assumption on reversible reactions by PEA, QSSA assumes that the production and consumption rates of some species (saying, the first $B$ species) are equal, so that they will stay in a quasi-steady state after a relatively short period of time \cite{gorban2005invariant}. We still adopt the same notation, a small parameter $0<\epsilon\ll 1$ to characterize the difference in the relaxation time between the first $B$ species and the remaining $(N-B)$ species. Therefore, the ODEs in \eqref{massactioneq} are rewritten as   
\begin{subequations}
\label{QSSA}
\begin{align}
\frac{{d}}{{dt}}c_k(t) &= 
-\frac{1}{\epsilon}\sum\limits_{i = 1}^M {\nu _{ik}} \left( {R_i ^ +  (\vec{c}) - R_i ^ -  (\vec{c})} \right),\;\;\;k = 1,2,...,B, \label{QSSA1}\\ 
\frac{{d}}{{dt}}c_k(t) &= 
-\sum\limits_{i=1}^M {\nu _{ik}} \left( {R_i ^ +  (\vec{c}) - R_i ^ -  (\vec{c})} \right),\;\;\;k = B+1,B+2,...,N. \label{QSSA2}
 \end{align}
\end{subequations}
By using QSSA, $B$ algebraic equations are obtained, and the original mass-action equations in \eqref{massactioneq} are reduced to
\begin{subequations}
\label{reQSSA}
\begin{align}
&\sum\limits_{i = 1}^M \nu _{ik} \left( {R_i ^ +  (\vec{c}) - R_i ^ -  (\vec{c})} \right) = 0,\;\;\;k = 1,2,...,B, \\ 
&\frac{{d}}{{dt}}c_k(t)=-\sum\limits_{i = 1}^M \nu _{ik} \left( {R_i ^ +  (\vec{c}) - R_i ^ -  (\vec{c})} \right),\quad c_k(0)=c_k^0, \quad k = B + 1,B + 2,...,N.
  \end{align}
\end{subequations}
Denote $f_k(\vec{c})=\sum\limits_{i = 1}^M \nu _{ik} \left( {R_i ^ +  (\vec{c}) - R_i ^ -  (\vec{c})} \right)$ ($k = 1,2,...,B$) and suppose its Jacobi matrix $[(\frac{\partial f_k(\vec{c})}{\partial c_j})]_{B\times B}$ is nondegenerate. Then according to the implicit function theorem, the first $B$ concentrations could be explicitly expressed as continuous and differentiable functions of the remaining $(N-B)$ concentrations, i.e. $c_k=c_k(c_{B+1},\cdots,c_N)$ for $k=1,2,\cdots, B$.

Alternatively, we may replace $B$ algebraic equations by the corresponding differential equations $dc_k(t)/dt=0$ with the original equilibrium values as $c_k^e$, which means 
\begin{equation}
\label{reQSSA-2}
\begin{array}{l}
 c_k(t) \equiv c_k^e,\;\;\;k = 1,2,...,B, \\ 
 \frac{{d}}{{dt}}c_k(t)=-\sum\limits_{i = 1}^M \nu _{ik} \left( {R_i ^ +  (\vec{c}) - R_i ^ -  (\vec{c})} \right),\quad c_k(0)=c_k^0, \quad k = B + 1,B + 2,...,N.
 \end{array}
\end{equation}
Note the equilibrium values are taken for the first $B$ species in order to keep the long-time consistency between the reduced model and the full model. Again, we call the reduced model in Eq. \eqref{reQSSA} as QSSA-1 and in Eq. \eqref{reQSSA-2} as QSSA-2.

\section{Non-equilibrium thermodynamics for reduced models of chemical reactions}
\label{ther reduced}

\subsection{Thermodynamics for full chemical reactions}
\label{noneq ther}
Chemical thermodynamics furnishes the fundamental principles of energy and forces within the realm of chemical species and their changes \cite{qian2021stochastic}. 
As to closed chemical reaction networks, their basic thermodynamic properties not only include the laws of thermodynamics but also include the equilibrium state, entropy, internal energy, free energy, entropy flow, entropy production, free energy dissipation, as well as their respect dependence on state variables. In the present work, we focus on the entropy and free energy functions, as well as their time derivatives, and decomposition. 

The core of non-equilibrium thermodynamics is an appropriate convex entropy function $Ent=Ent(t)$, from which the generalized Gibbs relation of entropy evolution follows. 
With respect to the chemical mass-action equations, the entropy can be chosen as 
\begin{equation}
\label{entropy}
Ent(t)=  - \sum\limits_{k = 1}^N {(c_k \ln } c_k  - c_k ).
\end{equation}
Taking time derivative of the entropy function and substituting the mass-action equation \eqref{massactioneq}, we have:
\begin{subequations}
\begin{align}
&\frac{d}{{dt}}Ent(t) = J^f(t)  + epr(t),\\
&J^f(t)=-\sum\limits_{i = 1}^M {(R_i ^ +  (\vec{c}) - R_i ^ -  (\vec{c}))\ln \frac{{\kappa _i ^ +  }}{{\kappa _i ^ -  }}} , 
\quad
epr(t)=
\sum\limits_{i = 1}^M {(R_i ^ +  (\vec{c}) - R_i ^ -  (\vec{c}))\ln } \frac{{R_i ^ +  (\vec{c})}}{{R_i ^ -  (\vec{c})}}\geq0
,
\end{align}
\end{subequations}
where the entropy flow rate $J^f(t)$ and entropy production rate $epr(t)$ are separated according to their different physical origins and mathematical properties. The entropy flow rate $J^f(t)$ represents the entropy flow into the system from its environment due to the exchange of matter and energy, while the entropy production rate $epr(t)$ is the entropy change in the system induced by irreversible processes, per second. The entropy production rate is always non-negative, and $epr(t)=0$ if and only if $R_i ^ +  (\vec{c^e}) = R_i ^ -  (\vec{c^e})$ for each $i$, or equivalently, the system is in equilibrium,  based on the second law of thermodynamics. 

Besides the entropy, we proceed to introduce the macroscopic free energy function $F(t)$ as
\begin{equation}
\label{free}
F(t)=\sum_{k=1}^N \left(c_k \ln\frac{c_k}{c_k^e} - c_k + c_k^e \right) \ge 0,
\end{equation}
which is non-negative, and it becomes zero if and only if the system reaches  the steady state, $\vec{c}=\vec{c^e}$.  
Correspondingly, the free energy dissipation rate is defined as the negative value of the time change rate of $F(t)$,
\begin{align}
f_d(t)  
&=  - \frac{dF(t)}{dt}
= \sum\limits_{i = 1}^M {(R_i ^ +  (\vec{c}) - R_i ^ -  (\vec{c}))\ln } \frac{{R_i ^ +  (\vec{c}) R_i^-(\vec{c^e}) }}{{R_i ^ -  (\vec{c})R_i ^ +  (\vec{c^e})}} \ge 0,
\end{align}
which is also non-negative, reflecting that the free energy is non-increasing along the trajectories of the mass-action equations. 
The free energy dissipation rate becomes zero, $f_d(t)=0$, if and only if when the steady state is reached. Thus, $F(t)$ serves as a Lyapunov function for the system of chemical mass-action equations, and the steady state is locally asymptotically stable. For a mass-action system with detailed balance condition, $R_i^+(\vec{c^e})=R_i^-(\vec{c^e})$ for each $i$, it is easily observed that $f_d(t)=epr(t)$ for the closed full system. However, this relation does not hold for open systems.   

We mention the above formulation holds for the full system before reduction. In the next part, we will look into the thermodynamic properties of the reduced system after applying PEA or QSSA, and especially focus on the correspondence between the reduced model and the original full model from the thermodynamic perspective.

\subsection{Thermodynamics for reduced models of chemical reactions by PEA}
\label{PEA thermodynamics}
\subsubsection{Thermodynamics for reduced PEA-1 model} 

With respect to the entropy function in Eq. \eqref{entropy}, the species are separated into two sets, consisting of the first $V$ species $\{c_k\}_{k=1}^{V}$ and the remaining ones $\{c_k\}_{k=V+1}^{N}$. By inserting the reduced PEA-1 model in Eq. \eqref{PEA1and2}, we arrive at 
\begin{eqnarray*}
\frac{d}{{dt}}\overline{Ent}(t)
&=& -\sum_{k=1}^V\frac{dc_k}{dt}\ln c_k -\sum_{k=V+1}^N\frac{dc_k}{dt}\ln c_k \nonumber\\
&=&-\frac{d}{dt}\left(\exp[\vec{b}_1^TA(A^TA)^{-1}]\right)(A^TA)^{-1}A^T\vec{b}_1
+\sum_{k=V+1}^N\sum\limits_{i = W + 1}^M {\nu _{ik}} \left( {R_i ^ +  (\vec{c}) - R_i ^ -  (\vec{c})} \right)\ln c_k\nonumber\\
&=&-\frac{d}{dt}\left(\exp[\vec{b}_1^TA(A^TA)^{-1}]\right)(A^TA)^{-1}A^T\vec{b}_1
+\sum\limits_{i = W + 1}^M  \left( {R_i ^ +  (\vec{c}) - R_i ^ -  (\vec{c})} \right)\ln\prod_{k=V+1}^N c_k^{\nu_{ik}}\nonumber\\
&=&-\frac{d}{dt}\left(\exp[\vec{b}_1^TA(A^TA)^{-1}]\right)(A^TA)^{-1}A^T\vec{b}_1-\sum\limits_{i = W + 1}^M  \left( {R_i ^ +  (\vec{c}) - R_i ^ -  (\vec{c})} \right) \vec{a}_i (A^TA)^{-1}A^T\vec{b}_1\nonumber\\
&&-\sum\limits_{i = W + 1}^M  \left( {R_i ^ +  (\vec{c}) - R_i ^ -  (\vec{c})} \right)\ln\frac{k_i^+}{k_i^-}
+\sum\limits_{i = W + 1}^M  \left( {R_i ^ +  (\vec{c}) - R_i ^ -  (\vec{c})} \right)\ln \frac{{R_i ^ +  (\vec{c})}}{{R_i ^ -  (\vec{c})}},
\end{eqnarray*}
where the matrix $A=(\nu_{ij})_{W\times V}$, vectors $\vec{a}_i=(\nu_{i1},\cdots,\nu_{iV})$ and $\vec{b}_1=(\ln (\kappa _1 ^ -/\kappa _1 ^ +)-\sum_{j=V+1}^N \nu_{1j}\ln c_j,\cdots, \ln (\kappa _W ^ -/\kappa _W ^ +)-\sum_{j=V+1}^N \nu_{Wj}\ln c_j)^T$. The over-line $\overline{\mathcal{C}}$ is utilized to denote the PEA-reduced results throughout the text. 

Based on the above formula, it is recognized that the entropy flow rate and entropy production rate are separately given by
\begin{eqnarray}
\label{PEA1:Jf}
\overline{J^f}(t)&=&-\frac{d}{dt}\left(\exp[\vec{b}_1^TA(A^TA)^{-1}]\right)(A^TA)^{-1}A^T\vec{b}_1-\sum\limits_{i = W + 1}^M  \left( {R_i ^ +  (\vec{c}) - R_i ^ -  (\vec{c})} \right)\vec{a}_i(A^TA)^{-1}A^T\vec{b}_1\nonumber\\
&&-\sum\limits_{i = W+1}^M {(R_i ^ +  (\vec{c}) - R_i ^ -  (\vec{c}))\ln \frac{{\kappa _i ^ +  }}{{\kappa _i ^ -  }}},\\
\overline{epr}(t)
&=&\sum\limits_{i = W+1}^M {(R_i ^ +  (\vec{c}) - R_i ^ -  (\vec{c}))\ln } \frac{{R_i ^ +  (\vec{c})}}{{R_i ^ -  (\vec{c})}}\geq0.
\end{eqnarray}
The latter is non-negative in accordance with the second law of thermodynamics. At the same time, by assuming that the detailed balance condition holds, the free energy function is still non-negative, $\overline{F}(t) =\sum_{k=1}^N (c_k \ln\frac{c_k}{c_k^e} - c_k + c_k^e) \geq 0$. However, the free energy dissipation rate for the PEA-1 model no longer has a definite sign, since
\begin{eqnarray}
\overline{f_d}(t)&=&-\frac{d\overline{F}}{{dt}}
= -\sum_{k=1}^N\frac{dc_k}{dt}\ln \frac{c_k}{c_k^e}\nonumber\\
&=&
\frac{d}{dt}\left(\exp[\vec{b}_1^TA(A^TA)^{-1}]\right)(A^TA)^{-1}A^T\vec{b}_2+\sum\limits_{i = W + 1}^M \left( {R_i ^ +  (\vec{c}) - R_i ^ -  (\vec{c})} \right)\ln \prod_{k=V+1}^N \left(\frac{c_k}{c_k^e}\right)^{\nu_{ik}},
\end{eqnarray}
where vector $\vec{b}_2=(\sum_{j=V+1}^N \nu_{1j}\ln (c_j/c_j^e),\cdots, \sum_{j=V+1}^N \nu_{Wj}\ln (c_j/c_j^e))^T$. The loss of non-negativity in the free energy dissipation rate in PEA-1 could be attributed to the breakdown of the mass-action law during the adoption of algebraic relations.

\subsubsection{Thermodynamics for reduced PEA-2 model}

Similarly, by inserting Eq. \eqref{PEA4} into the entropy change rate, we can construct the thermodynamics for the reduced PEA-2 model, i.e.
\begin{eqnarray*}
\frac{d}{{dt}}\overline{Ent}(t)
&=& \sum_{k=1}^N\sum\limits_{i = W + 1}^M {\nu _{ik}} \left( {R_i ^ +  (\vec{c}) - R_i ^ -  (\vec{c})} \right)\ln c_k=\sum\limits_{i = W + 1}^M \left( {R_i ^ +  (\vec{c}) - R_i ^ -  (\vec{c})} \right)\ln \prod_{k=1}^N c_k^{\nu_{ik}}\nonumber\\
&=&-\sum\limits_{i = W + 1}^M \left( {R_i ^ +  (\vec{c}) - R_i ^ -  (\vec{c})} \right)\ln \frac{\kappa _i ^ +}{\kappa _i ^ -}+\sum\limits_{i = W + 1}^M \left( {R_i ^ +  (\vec{c}) - R_i ^ -  (\vec{c})} \right)\ln \frac{R_i ^ +  (\vec{c})}{R_i ^ -  (\vec{c})},
\end{eqnarray*}
which shows that the entropy flow rate and entropy production rate are given by
\begin{eqnarray}
\overline{J^f}(t)&=&-\sum\limits_{i = W+1}^M {(R_i ^ +  (\vec{c}) - R_i ^ -  (\vec{c}))\ln \frac{{\kappa _i ^ +  }}{{\kappa _i ^ -  }}} , \\
\overline{epr}(t)
&=&\sum\limits_{i = W+1}^M {(R_i ^ +  (\vec{c}) - R_i ^ -  (\vec{c}))\ln } \frac{{R_i ^ +  (\vec{c})}}{{R_i ^ -  (\vec{c})}}\geq0.
\end{eqnarray}
Meanwhile, the free energy function is $\overline{F}(t) \geq 0$, and its dissipation rate reads
\begin{eqnarray}
\overline{f_d}(t)&=&
\sum_{k=1}^N\sum\limits_{i = W + 1}^M {\nu _{ik}} \left( {R_i ^ +  (\vec{c}) - R_i ^ -  (\vec{c})} \right)\ln \frac{c_k}{c_k^e}=\sum\limits_{i = W + 1}^M \left( {R_i ^ +  (\vec{c}) - R_i ^ -  (\vec{c})} \right)\ln \prod_{k=1}^N \left(\frac{c_k}{c_k^e}\right)^{\nu_{ik}}\nonumber\\
&=&\sum\limits_{i = W + 1}^M \left( {R_i ^ +  (\vec{c}) - R_i ^ -  (\vec{c})} \right)\ln \frac{R_i ^ +  (\vec{c})R_i ^ -  (\vec{c^e})}{R_i ^ -  (\vec{c})R_i ^ +  (\vec{c^e})}\geq0,
\end{eqnarray}
which is always non-negative. 

A direct comparison on the thermodynamics of PEA-1, PEA-2 and the full model shows that:  (1)  formulas for the entropy flow rate, the entropy production rate and the free energy dissipation rate of the reduced model PEA-2 are all a part, the last $(M-W)$ reactions, of their corresponding ones in the full model, by just neglecting the contributions from the fast $W$ reactions. Furthermore, its entropy production rate exactly equals to the free energy dissipation rate, both of which are non-negative. (2) The entropy production rate for PEA-1 is always non-negative, while its free energy dissipation rate is not. This fact could be attributed to the breakdown of the law of mass-action in PEA-1, during the adoption of algebraic relations for model reduction. (3) Despite of their similar forms, values of the entropy production rate for PEA-1, PEA-2 are not necessarily the same, since the concentrations of species $\vec{c}$ in the two models may not coincide with each other. 

These results will be further verified through numerical simulations of enzymatic reactions in Sect. \ref{MM PEA therm}. In conclusion, we claim that the PEA-2 method can well preserve the essential thermodynamic properties of the full model during model reduction, in particular the ``closed'' nature of the reaction system and the law of mass action.

\subsection{Thermodynamics for reduced models of chemical reactions by QSSA}
\label{QSSA thermodynamics}
\subsubsection{Thermodynamics for reduced QSSA-1 model}

Different from the PEA method, we can not get an explicit closed expression of the entropy production rate or the entropy flow rate for the QSSA-1 model. However, as we claimed in the Sect. \ref{Quasi steady state approximation}, in QSSA the first $B$ concentrations could be explicitly expressed as continuous and differentiable functions of the remaining $(N-B)$ concentrations, i.e. $c_k=c_k(c_{B+1},\cdots,c_N)$ for $k=1,2,\cdots, B$. As a consequence, once the reduced equations in Eq. \eqref{reQSSA} are solved, either analytically or numerically, we can substitute the solutions into the thermodynamic relations presented in Sec. \ref{noneq ther} and obtain the corresponding reduced ones. This strategy is illustrated as follows in a formal way, and verified by enzymatic reactions in the next section. 

Thus for QSSA-1 , its entropy change rate reads
\begin{eqnarray*}
\frac{d}{{dt}}\widetilde{Ent}(t)
&=&-\sum_{k=1}^B \frac{dc_k}{dt}\ln c_k +\sum_{k=B+1}^N\sum\limits_{i = 1}^M {\nu _{ik}} \left( {R_i ^ +  (\vec{c}) - R_i ^ -  (\vec{c})} \right)\ln c_k\nonumber\\
&=&-\sum_{k=1}^B \sum_{j=B+1}^N\frac{\partial c_k}{\partial c_j}\frac{dc_j}{dt}\ln c_k +\sum_{k=B+1}^N\sum\limits_{i = 1}^M {\nu _{ik}} \left( {R_i ^ +  (\vec{c}) - R_i ^ -  (\vec{c})} \right)\ln c_k\nonumber\\
&=&\sum_{k=1}^B\sum_{j=B+1}^N\frac{\partial c_k}{\partial c_j}\sum\limits_{i = 1}^M {\nu _{ij}} \left( {R_i ^ +  (\vec{c}) - R_i ^ -  (\vec{c})} \right)\ln c_k+\sum_{k=B+1}^N\sum\limits_{i = 1}^M {\nu _{ik}} \left( {R_i ^ +  (\vec{c}) - R_i ^ -  (\vec{c})} \right)\ln c_k\nonumber\\
&=&\sum_{k=B+1}^N\sum\limits_{i = 1}^M {\nu _{ik}} \left( {R_i ^ +  (\vec{c}) - R_i ^ -  (\vec{c})} \right)\left(\sum_{j=1}^B\frac{\partial c_j}{\partial c_k}\ln c_j+\ln c_k\right)=\sum_{k=B+1}^N\sum\limits_{i = 1}^M {\nu _{ik}} \left( {R_i ^ +  (\vec{c}) - R_i ^ -  (\vec{c})} \right)\ln \tilde{c}_k\nonumber\\
&=&-\sum\limits_{i = 1}^M \left( {R_i ^ +  (\vec{c}) - R_i ^ -  (\vec{c})} \right)\ln \prod_{k=1}^B c_k^{\nu_{ik}}-\sum\limits_{i = 1}^M \left( {R_i ^ +  (\vec{c}) - R_i ^ -  (\vec{c})} \right)\ln \frac{\kappa _i ^ +}{\kappa _i ^ -}+\sum_{k=B+1}^N\sum\limits_{i = 1}^M {\nu _{ik}} \left( {R_i ^ +  (\vec{c}) - R_i ^ -  (\vec{c})} \right)\ln \frac{\tilde{c}_k}{c_k}\nonumber\\
&&+\sum\limits_{i = 1}^M \left( {R_i ^ +  (\vec{c}) - R_i ^ -  (\vec{c})} \right)\ln \frac{R_i ^ +  (\vec{c})}{R_i ^ -  (\vec{c})},
\end{eqnarray*}
where $\tilde{c}_k=c_k\prod_{j=1}^B \exp(\partial c_j/\partial c_k\ln c_j)$ for $k=B+1,\cdots,N$. The tilde $\widetilde{\mathcal{C}}$ is utilized to denote the QSSA-reduced results. 
A direct decomposition gives the entropy flow rate and entropy production rate, i.e.
\begin{eqnarray}
\widetilde{J^f}(t)&=&-\sum\limits_{i = 1}^M \left( {R_i ^ +  (\vec{c}) - R_i ^ -  (\vec{c})} \right)\ln \prod_{k=1}^B c_k^{\nu_{ik}}-\sum\limits_{i = 1}^M \left( {R_i ^ +  (\vec{c}) - R_i ^ -  (\vec{c})} \right)\ln \frac{\kappa _i ^ +}{\kappa _i ^ -}+\sum_{k=B+1}^N\sum\limits_{i = 1}^M {\nu _{ik}} \left( {R_i ^ +  (\vec{c}) - R_i ^ -  (\vec{c})} \right)\ln \frac{\tilde{c}_k}{c_k} , \\
\widetilde{epr}(t)
&=&\sum\limits_{i = 1}^M {(R_i ^ +  (\vec{c}) - R_i ^ -  (\vec{c}))\ln } \frac{{R_i ^ +  (\vec{c})}}{{R_i ^ -  (\vec{c})}}\geq0.
\end{eqnarray}
At the same time, the free energy function remains non-negative, $\widetilde{F}(t) =\sum_{k=1}^N (c_k \ln\frac{c_k}{c_k^e} - c_k + c_k^e) \geq 0$. Nevertheless, the free energy dissipation rate reads
\begin{eqnarray}
\widetilde{f_d}(t)&=&-\sum_{k=1}^B \frac{dc_k}{dt}\ln \frac{c_k}{c_k^e}
+\sum_{k=B+1}^N\sum\limits_{i =1}^M {\nu _{ik}} \left( {R_i ^ +  (\vec{c}) - R_i ^ -  (\vec{c})} \right)\ln \frac{c_k}{c_k^e}\nonumber\\
&=&\sum_{k=B+1}^N\sum\limits_{i = 1}^M {\nu _{ik}} \left( {R_i ^ +  (\vec{c}) - R_i ^ -  (\vec{c})} \right) \left(\sum_{j=1}^B\frac{\partial c_j}{\partial c_k}\ln \frac{c_j}{c_j^e}+\ln \frac{c_k}{c_k^e}\right),
\end{eqnarray}
whose sign is undetermined, similar to the PEA-1.

\subsubsection{Thermodynamics for reduced QSSA-2 model}

In contrast, the thermodynamics for the reduced QSSA-2 model is far simpler. 
With respect to the state variables $\vec{c}(t)=(c_1^e, \cdots, c_B^e, c_{B+1}(t), \cdots, c_N(t))^T$, the entropy function becomes $\widetilde{Ent}(t)$. The entropy change rate reads
\begin{eqnarray*}
\frac{d}{{dt}}\widetilde{Ent}(t)
&=& \sum_{k=B+1}^N\sum\limits_{i = 1}^M {\nu _{ik}} \left( {R_i ^ +  (\vec{c}) - R_i ^ -  (\vec{c})} \right)\ln c_k=\sum\limits_{i = 1}^M \left( {R_i ^ +  (\vec{c}) - R_i ^ -  (\vec{c})} \right)\ln \prod_{k=B+1}^N c_k^{\nu_{ik}}\\
&=&-\sum\limits_{i = 1}^M \left( {R_i ^ +  (\vec{c}) - R_i ^ -  (\vec{c})} \right)\ln \prod_{k=1}^B (c_k^e)^{\nu_{ik}}-\sum\limits_{i = 1}^M \left( {R_i ^ +  (\vec{c}) - R_i ^ -  (\vec{c})} \right)\ln \frac{\kappa _i ^ +}{\kappa _i ^ -}\\
&&+\sum\limits_{i = 1}^M \left( {R_i ^ +  (\vec{c}) - R_i ^ -  (\vec{c})} \right)\ln \frac{R_i ^ +  (\vec{c})}{R_i ^ -  (\vec{c})},
\end{eqnarray*}
whose two components, the entropy flow rate and entropy production rate, are given through the following simple relations,
\begin{eqnarray}
\widetilde{J^f}(t)&=&-\sum\limits_{i = 1}^M \left( {R_i ^ +  (\vec{c}) - R_i ^ -  (\vec{c})} \right)\ln \prod_{k=1}^B (c_k^e)^{\nu_{ik}}-\sum\limits_{i = 1}^M \left( {R_i ^ +  (\vec{c}) - R_i ^ -  (\vec{c})} \right)\ln \frac{\kappa _i ^ +}{\kappa _i ^ -} , \\
\widetilde{epr}(t)
&=&\sum\limits_{i = 1}^M {(R_i ^ +  (\vec{c}) - R_i ^ -  (\vec{c}))\ln } \frac{{R_i ^ +  (\vec{c})}}{{R_i ^ -  (\vec{c})}}\geq0.
\end{eqnarray}
More importantly, the free energy function $\widetilde{F}(t) \geq 0$ and the dissipation rate for QSSA-2 take the same form as those of the full system and are always non-negative too, 
\begin{eqnarray}
\label{QSSA2:fd}
\widetilde{f_d}(t)&=&
\sum_{k=B+1}^N\sum\limits_{i =1}^M {\nu _{ik}} \left( {R_i ^ +  (\vec{c}) - R_i ^ -  (\vec{c})} \right)\ln \frac{c_k}{c_k^e}=\sum\limits_{i = 1}^M \left( {R_i ^ +  (\vec{c}) - R_i ^ -  (\vec{c})} \right)\ln \prod_{k=B+1}^N \left(\frac{c_k}{c_k^e}\right)^{\nu_{ik}}\nonumber\\
&=&\sum\limits_{i = 1}^M \left( {R_i ^ +  (\vec{c}) - R_i ^ -  (\vec{c})} \right)\ln \frac{R_i ^ +  (\vec{c})R_i ^ -  (\vec{c^e})}{R_i ^ -  (\vec{c})R_i ^ +  (\vec{c^e})}\geq0.
\end{eqnarray}

In summary, (1) the QSSA-2 method preserves the most basic thermodynamic relations as in the full model. In particular, the forms of entropy production rate and free energy dissipation rate of the QSSA-2 model keep the same as those of the full model, though their values can be different. (2) There is an additional term in the entropy flow rate of QSSA-2, which accounts for the contributions from the QSSA approximation on $(c_1,\cdots,c_B)$. (3) The loss of non-negativity in the free energy dissipation rate of QSSA-1 could be attributed to the usage of algebraic relations, which violates the law of mass-action and makes the system from closed to open. (4) Formally, the entropy production rates of QSSA-1 and QSSA-2 contain the contributions of all $M$ reactions of the full system, in contrast to the entropy production rates of PEA-1 and PEA-2 including only $(M-W)$ reactions. This fact reflects an intrinsic difference between the PEA and QSSA. Generally speaking, the PEA method removes the fast reactions, while the QSSA makes a reduction on the fast variables.

The dynamical and thermodynamic features of the full model and reduced models by either PEA or QSSA for general chemical reactions discussed  above are compared and summarized in Table \ref{table1}. Those features can be applied to MM reactions in the next section. 

\begin{table}[h]
    \centering
    \includegraphics[width=1.0\linewidth]{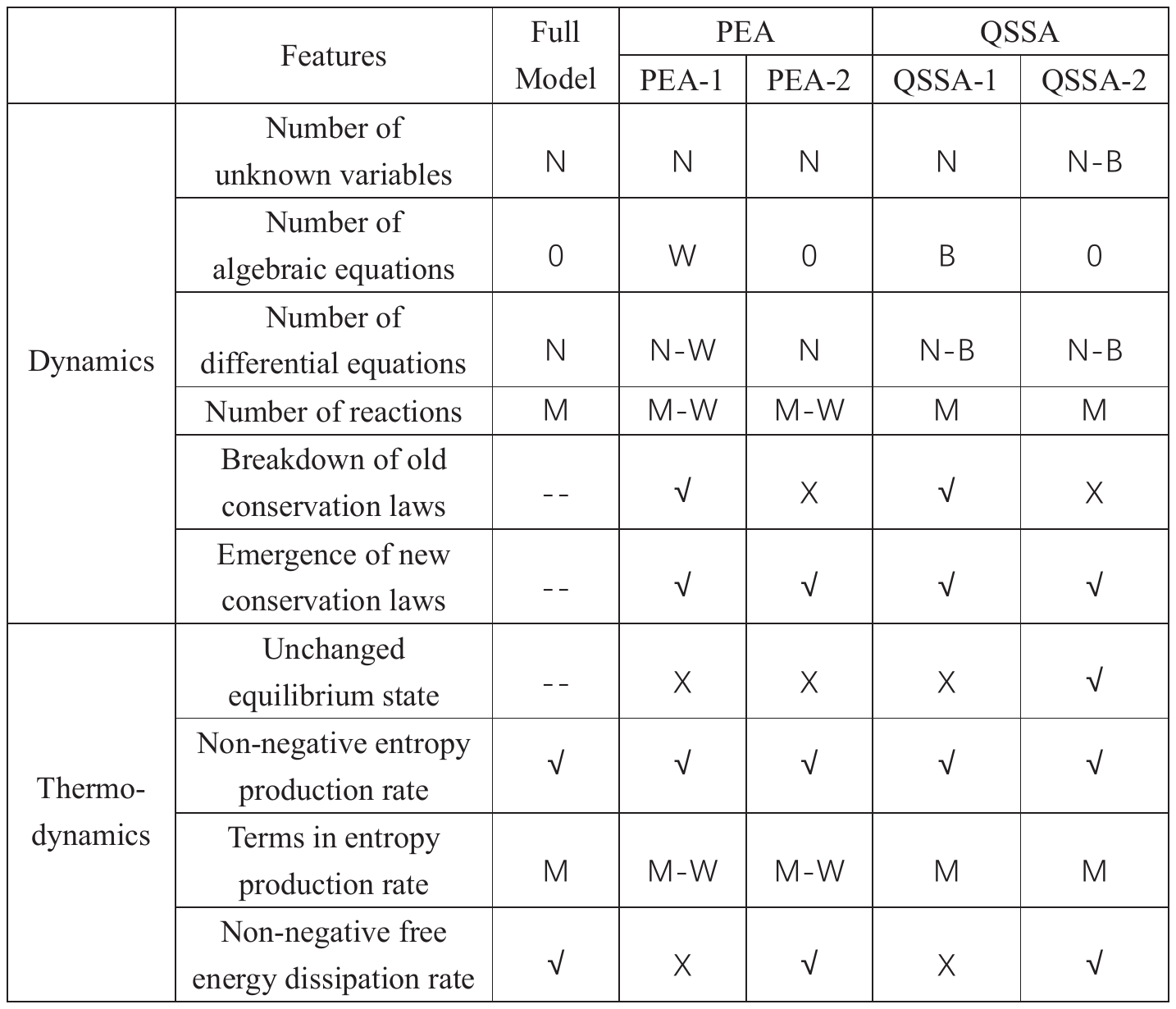}
    \caption{Comparison on key features of the full model and reduced models by either PEA or QSSA for general reversible chemical reactions discussed in Sect. \ref{ther reduced}. As to dynamical features of the models reduced by PEA-1, the unknown variables are the original ones $\{c_k\}_{k=1}^N$, governed by both $W$ algebraic equations and $(N-W)$ ODEs, in Eq. \eqref{PEA1and2}. Once adopted PEA, the $W$ fast reactions are eliminated, and $(M-W)$ slow ones are remained. The old conservation laws may break down due to the replacement of $W$ ODEs by the corresponding algebraic relations. Meanwhile, new conservation laws may emerge too. 
    As to thermodynamic features of PEA-1 models, the equilibrium state of the reduced model is generally different from the one of the full model (see e.g. discussions in Sect. \ref{conservation-MM-reaction}). The entropy production rate, contributed by $(M-W)$ slow reactions, is preserved to be non-negative, while the free energy dissipation rate no longer remains non-negative. 
    The illustration of other three reduced models follows analogously.}
    \label{table1}
\end{table}

\section{Application to Michaelis-Menten reactions}
\label{Closed Michaelis-Menten reactions}
In this section, we will apply the general formulation of PEA and QSSA thermodynamics to the famous Michaelis-Menten reactions (MM) as an illustration, theoretically and numerically. The MM reactions are based on the enzyme-substrate binding mechanism and have been proposed to explain the dramatic catalytic effects of enzymes on the rates of chemical reactions \cite{johnson2011original}. Here we consider a closed system of MM reactions where all species are confined in a finite volume without exchange of both material and energy. 
Given the fact that the substrate $S$ is catalyzed by enzyme $E$ to get product $P$, the MM reactions assume the existence of an intermediate complex $C \equiv SE$: 
\begin{equation}
\label{MM}
\underbrace{ S{\text{ + }}E \xrightleftharpoons[{\kappa_1^-}]{{\kappa_1^+}}
}_{\text{Reaction}~1}  C \underbrace{ \xrightleftharpoons[{\kappa_2^-}]{{\kappa_2^+}} P{\text{ + }}E }_{\text{Reaction}~2},
\end{equation}
where two reversible reactions are considered for the convenience of thermodynamic analysis, the parameters $\kappa_1^+>0$ and $\kappa_1^->0$ (or, $\kappa_2^+>0$ and $\kappa_2^-\geq 0$) are rate constants of the forward and backward reactions for substrate binding (or, substrate conversion) respectively. Note that, $\kappa_2^-$ is usually assumed to be small compared to $\kappa_2^+$, and when $\kappa_2^- \rightarrow 0$, the above mechanism degenerates into the classic MM model. 

According to the mass-action law of two reversible elementary reactions in \eqref{MM}, we have 
\begin{subequations}
\label{mass-action-MM}
\begin{align}
 \frac{{d[S]}}{{dt}}&=  - \kappa _1 ^ +  [S][E] + \kappa _1 ^ -  [C], \\ 
 \frac{{d[E]}}{{dt}}&=  - \kappa _1 ^ +  [S][E] + (\kappa _1 ^ -   + \kappa _2 ^ +  )[C] - \kappa _2 ^ -  [P][E], \\ 
 \frac{{d[C]}}{{dt}}&= \kappa _1 ^ +  [S][E] - (\kappa _1 ^ -   + \kappa _2 ^ +  )[C] + \kappa _2 ^ -  [P][E], \\ 
 \frac{{d[P]}}{{dt}}&= \kappa_2^+  [C] - \kappa_2^-  [P][E],
 \end{align}
\end{subequations}
where the initial concentrations of the substrate, enzyme, complex, and product are assumed to be $[S]_0>0, [E]_0>0, [C]_0\geq 0, [P]_0\geq 0$ respectively. Combining Eq. \eqref{mass-action-MM} with the above initial condition, two relations can be obtained as 
\begin{equation}
\text{(Conservation law 1)}~ [E] + [C] = [E]_0 + [C]_0, \quad \text{(Conservation law 2)}~ [S] + [C] + [P] = [S]_0 + [C]_0 + [P]_0, 
\end{equation}
which are also known as the conservation law of enzyme and of substrate respectively for closed MM reactions \eqref{MM}. As a consequence, the system in \eqref{mass-action-MM} possesses only two independent variables, saying $[S]$ and $[C]$. 
Note that the above two conservation laws hold for the full system, which may get broken after the adoption of PEA or QSSA methods. We will go to details later.  

For MM reactions, the number of species and reactions are $N=4$ and $M=2$ respectively. 
Denote $\vec{c}(t) \equiv (c_1 ,c_2 ,c_3 ,c_4 )^T = ([S],[E],[C],[P])^T$. 
There exists a unique positive equilibrium state, $\vec{c^e} \equiv (c^e_1, c^e_2, c^e_3, c^e_4)^T>0$ for reversible MM reactions in \eqref{mass-action-MM}, which is illustrated in detail as follows. 

After some manipulations of Eq. \eqref{mass-action-MM}, the equilibrium state of the enzyme $c^e_2=[E]^e$ satisfies the following algebraic equation of second order when $\kappa_2^- >0$,
\begin{equation}
\label{quad}
    \underbrace{ \kappa_1^+\kappa_2^- (c^e_2) ^2 + \left[\kappa_1^+\kappa_2^+ +  \kappa_1^-\kappa_2^- + \kappa_1^+\kappa_2^-\left([S]_0 + [P]_0 - [E]_0\right)\right]c^e_2 - (\kappa_1^+\kappa_2^+ +  \kappa_1^-\kappa_2^-)([E]_0+[C]_0) }_{\text{quadratic function}~g{(c_2^e)}}=0, 
\end{equation}
from which the two solutions are derived as 
\begin{align}
    &c^e_2=\frac{-\left[\kappa_1^+\kappa_2^+ +  \kappa_1^-\kappa_2^- + \kappa_1^+\kappa_2^-\left([S]_0 +[P]_0 - [E]_0\right)\right] \pm \sqrt{\Delta}}{2\kappa_1^+\kappa_2^-}
    , \\ 
    &\Delta = \left[\kappa_1^+\kappa_2^+ +  \kappa_1^-\kappa_2^- + \kappa_1^+\kappa_2^-\left([S]_0 +[P]_0 - [E]_0\right)\right]^2 + 4 \kappa_1^+\kappa_2^- (\kappa_1^+\kappa_2^+ +  \kappa_1^-\kappa_2^-)([E]_0+[C]_0) > 0.
\end{align}
Denote the quadratic function on the left-hand side of Eq. \eqref{quad} as $g(c^e_2)$, which satisfies that: (i) the coefficient of the squared term is positive, $\kappa_1^+\kappa_2^- >0$; (ii) the values $g(c^e_2=0)=- (\kappa_1^+\kappa_2^+ +  \kappa_1^-\kappa_2^-)([E]_0+[C]_0) <0$ and $g(c^e_2=[E]_0+[C]_0)=\kappa_1^+\kappa_2^-([E]_0+[C]_0)([S]_0+[C]_0+[P]_0)  >0$. As a result, there exists exactly one solution such that $0<c^e_2<[E]_0+[C]_0$. The conservation of enzyme yields $0<c^e_3<[E]_0+[C]_0$. The relations $c^e_1 = \kappa_1^-c^e_3/(\kappa_1^+ c^e_2) >0$, $c^e_4/c^e_1=\kappa_1^+\kappa_2^+ / (\kappa_1^-\kappa_2^-)>0$, and $c^e_1 + c^e_3 + c^e_4 = [S]_0+[C]_0+[P]_0$ together yield that $0<c^e_1, c^e_3, c^e_4<[S]_0+[C]_0+[P]_0$. This completes the proof. 


\subsection{Non-equilibrium thermodynamics for the full MM reactions}
Choosing the entropy and free energy functions for the full MM reactions as
\begin{equation}
Ent(t)=  - \sum\limits_{k = 1}^4 {(c_k \ln } c_k  - c_k ), \quad
F(t)=\sum_{k=1}^4 \left(c_k \ln\frac{c_k}{c_k^e} - c_k + c_k^e \right),
\end{equation} 
we have the following thermodynamic quantities and their relations as 
\begin{subequations}
\label{MM_thermo}
\begin{align}
\frac{d}{{dt}}Ent(t)&= J^f(t)  + epr(t),\\
J^f(t)&=-(\kappa _1 ^ +  [S][E] - \kappa _1 ^ -  [C])\ln \frac{{\kappa _1 ^ +  }}{{\kappa _1 ^ -  }} - (\kappa _2 ^ +  [C] - \kappa _2 ^ -  [P][E])\ln \frac{{\kappa _2 ^ +  }}{{\kappa _2 ^ -  }}, \\
epr(t)&=(\kappa _1 ^ +  [S][E] - \kappa _1 ^ -  [C])\ln \frac{{\kappa _1 ^ +  [S][E]}}{{\kappa _1 ^ -  [C]}} + (\kappa _2 ^ +  [C] - \kappa _2 ^ -  [P][E])\ln \frac{{\kappa _2 ^ +  [C]}}{{\kappa _2 ^ -  [P][E]}} \ge 0,\\
f_d(t)&=-\frac{dF}{dt}
=(\kappa _1 ^ +  [S][E] - \kappa _1 ^ -  [C])\ln \frac{{[S][E][C]^e}}{{[S]^e[E]^e[C]}} + (\kappa _2 ^ +  [C] - \kappa _2 ^ -  [P][E])\ln \frac{{ [E]^e[C][P]^e }}{{[E][C]^e[P]}}.
\end{align}
\end{subequations}
According to the detailed balance condition, the relation $f_d(t)=epr(t)\ge 0$ holds for the full MM reactions.

\subsection{Thermodynamics for PEA-reduced MM reactions}
\label{MM PEA therm}
Here we assume the association and disassociation of the substrate and enzyme (Reaction 1) proceed in a time scale much shorter than Reaction 2, which means Reaction 1, $S+E \xrightleftharpoons[{\kappa_1^-}]{{\kappa_1^+}} C$, is considered to be fast. By  explicitly writing out the order $\epsilon$ of MM reactions, we have 
\begin{subequations}
\label{PEA-MM}
\begin{align*}
 \frac{{d[S]}}{{dt}}&= -\frac{1}{\epsilon} \left(\widehat{R_1^+}(\vec{c}) - \widehat{R_1^-}(\vec{c}) \right), \\ 
 \frac{{d[E]}}{{dt}}&=-\frac{1}{\epsilon}\left(\widehat{R_1^+}(\vec{c}) - \widehat{R_1^-}(\vec{c}) \right) +(R_2^+(\vec{c}) - R_2^-(\vec{c})), \\ 
 \frac{{d[C]}}{{dt}}&=\frac{1}{\epsilon}\left(\widehat{R_1^+}(\vec{c}) - \widehat{R_1^-}(\vec{c}) \right) -(R_2^+(\vec{c}) - R_2^-(\vec{c})), \\ 
 \frac{{d[P]}}{{dt}}&=R_2^+(\vec{c}) - R_2^-(\vec{c}),
 \end{align*}
\end{subequations}
where $\widehat{R_1^{\pm}}(\vec{c}) \equiv \epsilon R_1^{\pm}(\vec{c})$ are of the same order as ${R_2^{\pm}}(\vec{c})$, and $R_1^{+}(\vec{c})=\kappa _1 ^ +  [S][E], R_1^{-}(\vec{c})=\kappa_1^-[C]$, $R_2^{+}(\vec{c})=\kappa _2^+[C], R_2^{-}(\vec{c})=\kappa_2^-[P][E]$. Taking the limit $\epsilon \rightarrow 0$, we have the algebraic equation, called the PEA relation $R_1^{+}(\vec{c}) =  R_1^{-}(\vec{c})$, or equivalently, 
\begin{equation}
\label{R1pm}
\kappa _1 ^ +  [S][E] = \kappa_1^-  [C].
\end{equation}

In this stage, we have four reduced ODEs 
\begin{subequations}
\label{mass-action-MM2}
\begin{align}
 \frac{{d[S]}}{{dt}}&= 0, \\ 
 \frac{{d[E]}}{{dt}}&= \kappa _2 ^ +  [C] - \kappa _2 ^ -  [P][E], \\ 
 \frac{{d[C]}}{{dt}}&= -(\kappa _2 ^ +  [C] - \kappa _2 ^ -  [P][E]), \\ 
 \frac{{d[P]}}{{dt}}&= \kappa_2^+  [C] - \kappa_2^-  [P][E],
 \end{align}
\end{subequations}
for four variables with the corresponding initial values $([S]_0, [E]_0, [C]_0, [P]_0)^T$ and an extra PEA relation $R_1^{+}(\vec{c}) =  R_1^{-}(\vec{c})$. These five equations constitute an over-determined system for four unknowns. Therefore, we begin to discuss the reduced MM model based on two different approaches, including PEA-MM1 and PEA-MM2 (notations for PEA-reduced models of MM reactions by route 1 and route 2 respectively).

\subsubsection {Thermodynamics for PEA-MM1: Use the algebraic relation.}

As to this case, there is only one fast reaction, $W=1$, and the concentration $[S]$ is expressed by the other ones as $[S]=k_1^-[C]/(k_1^+[E])$, which indicates that $V=1$. Following the general theory established above, we neglect the ODE $d[S]/dt=0$ and obtain the PEA-reduced dynamics of the MM reactions in terms of the state variables $([E], [C], [P])^T$ as ${d[E]}/{dt}=-{d[C]}/{dt}={d[P]}/{dt}= \kappa_2^+  [C] - \kappa_2^-  [P][E]$, 
here the initial values are $([E]_0, [C]_0, [P]_0)^T$. The concentration $[S]=k_1^-[C]/(k_1^+[E])$ decouples with the state variables in ODEs and thus can be obtained by substitution of the solutions to above ODEs.  

We proceed to derive the entropy change rate of the PEA-MM1 mechanism by recalling that $\ln{[S]}=-\ln{\frac{\kappa _1 ^ + [E]}{\kappa_1^-  [C]}}$ and $\frac{d[S]}{dt}=-\frac{[S]}{[E]}\frac{d[E]}{dt}+\frac{[S]}{[C]}\frac{d[C]}{dt}=\frac{\kappa_1^-  [C]}{\kappa _1 ^ + [E]}(-\frac{1}{[E]}\frac{d[E]}{dt}+\frac{1}{[C]}\frac{d[C]}{dt})$,
\begin{align*}
\frac{d}{dt}\overline{Ent}(t)&=-\ln{[S]}\frac{d[S]}{dt} -\ln{[E]}\frac{d[E]}{dt} -\ln{[C]}\frac{d[C]}{dt} -\ln{[P]}\frac{d[P]}{dt}\\
&=\ln\left( \frac{\kappa _1 ^ + [E]}{\kappa_1^-  [C]} \right) \frac{\kappa_1^-  [C]}{\kappa _1 ^ + [E]} \left(-\frac{1}{[E]}\frac{d[E]}{dt}+\frac{1}{[C]}\frac{d[C]}{dt} \right) 
-\ln{[E]}\frac{d[E]}{dt} -\ln{[C]}\frac{d[C]}{dt} -\ln{[P]}\frac{d[P]}{dt}\\
&=(\kappa _2 ^ +  [C] - \kappa _2 ^ -  [P][E])\ln \left[ \left(\frac{{[C]}}{{[P][E]}} \right) 
\left( \frac{\kappa_1^-  [C]}{\kappa _1 ^ + [E]} \right)^{\left( \frac{\kappa_1^- ([E]_0+[C]_0)}{\kappa _1 ^ + [E]^2} \right) } 
\right],
\end{align*}
from which the entropy production rate and entropy flow rate are separated and obtained as 
\begin{align}
    \overline{epr}(t)&=(\kappa _2 ^ +  [C] - \kappa _2 ^ -  [P][E])\ln \frac{\kappa _2 ^ +  [C]}{\kappa _2 ^ -  [P][E]} \geq 0, \\
    \overline{J^f}(t)&=(\kappa _2 ^ +  [C] - \kappa _2 ^ -  [P][E])\ln \left[ \frac{\kappa _2 ^ -}{\kappa _2 ^ +}
\left( \frac{\kappa_1^-  [C]}{\kappa _1 ^ + [E]} \right)^{\left( \frac{\kappa_1^- ([E]_0+[C]_0)}{\kappa _1 ^ + [E]^2} \right) } 
\right].
\end{align}
Analogously, the free energy dissipation rate is calculated as 
\begin{align}
\label{fd-PEAMM1}
\overline{f_d}(t) 
&\equiv 
-\frac{d\overline{F}(t)}{dt}
=-\ln\left(\frac{[S]}{[S]^e}\right) \frac{d[S]}{dt} -\ln\left(\frac{[E]}{[E]^e}\right)\frac{d[E]}{dt} -\ln\left(\frac{[C]}{[C]^e}\right)\frac{d[C]}{dt} -\ln\left(\frac{[P]}{[P]^e}\right)\frac{d[P]}{dt}\nonumber\\
&=-\ln\left( \frac{[E]^e[C]}{[E][C]^e} \right) \frac{\kappa_1^-  [C]}{\kappa _1 ^ + [E]} \left(-\frac{1}{[E]}\frac{d[E]}{dt}+\frac{1}{[C]}\frac{d[C]}{dt} \right) 
-\ln\left(\frac{[E]}{[E]^e}\right)\frac{d[E]}{dt} -\ln\left(\frac{[C]}{[C]^e}\right)\frac{d[C]}{dt} -\ln\left(\frac{[P]}{[P]^e}\right)\frac{d[P]}{dt}\nonumber\\
&=(\kappa _2 ^ +  [C] - \kappa _2 ^ -  [P][E])\ln \left[ \left(\frac{{[E]^e[C][P]^e}}{{[E][C]^e[P]}} \right) 
\left( \frac{[E]^e[C]}{[E][C]^e} \right)^{\left( \frac{\kappa_1^- ([E]_0+[C]_0)}{\kappa _1 ^ + [E]^2} \right) } 
\right],
\end{align}
which has a undetermined sign. 

\subsubsection{Thermodynamics for PEA-MM2: Use no algebraic relation.}

The governing equations for PEA-MM2 are ODEs ${d[E]}/{dt}=-{d[C]}/{dt}={d[P]}/{dt}= \kappa_2^+  [C] - \kappa_2^-  [P][E]$ with the initial values $([E]_0, [C]_0, [P]_0)^T$ and the concentration of the substrate $[S](t)\equiv [S]^e$ by assigning its positive equilibrium value. 
By applying the thermodynamics of general reactions reduced via PEA-2 in Sect. \ref{PEA thermodynamics} to MM reactions, we have 
\begin{subequations}
\label{MM_thermo}
\begin{align}
\overline{J^f}(t)&=-(\kappa _2 ^ +  [C] - \kappa _2 ^ -  [P][E])\ln \frac{{\kappa _2 ^ +  }}{{\kappa _2 ^ -  }}, \\
\overline{epr}(t)&=(\kappa _2 ^ +  [C] - \kappa _2 ^ -  [P][E])\ln \frac{{\kappa _2 ^ +  [C]}}{{\kappa _2 ^ -  [P][E]}} \ge 0, \\
\overline{f_d}(t)&=(\kappa _2 ^ +  [C] - \kappa _2 ^ -  [P][E])\ln \frac{{[E]^e [C] [P]^e}}{{[E] [C]^e [P]}} \ge 0.
\end{align}
\end{subequations}
The above expressions show that the entropy production rate of the PEA-MM2 model is preserved to be non-negative, $\overline{epr}(t) \geq 0$, duo to the monotonicity of $\ln(\cdot)$ function; and $\overline{f_d}(t)=\overline{epr}(t)$. Remark that the variables in Eqs. \eqref{MM_thermo} are solutions to the PEA-MM2 model, although they are denoted with the same notations as the original dynamics for simplicity.

\subsubsection{Discussions on conservation laws}
\label{conservation-MM-reaction}

It is well-known that, during the procedure of model reduction, old conservation laws may be broken while new conservation laws may emerge. As a result, in our above studies on both PEA thermodynamics and QSSA thermodynamics, no conservation law is considered. This is essential for the validity of our general formulation. For a simple system like MM reactions, we can go a step further and examine the effect of conservation laws on the dynamics, especially the steady state, of reduced models as well as their thermodynamic aspect. 

Let's take the PEA method as an example. The last three formulas in Eq. \eqref{mass-action-MM2} state the existence of two conservation laws in both PEA reduced models, i.e. $[E]+[C]=Const$ and $[C]+[P]=Const$. It is clear that the first one coincides with the Conservation law 1 for the full system, while the second one is new. The emergence of new conservation laws brings complication to the PEA reduced MM reactions. For PEA-MM1, the steady state satisfies that 
\begin{equation*}
    [E]+[C]=[E]_0+[C]_0, ~[C]+[P]=[C]_0+[P]_0, ~ 
    \kappa _1 ^ +  [S][E] = \kappa_1^-  [C], ~
    \kappa_2^+  [C]       = \kappa_2^-  [P][E],
\end{equation*}
whose solution is usually different from the equilibrium state $\vec{c^e}$ of the full model. 
While for PEA-MM2, the steady state satisfies that 
\begin{equation*}
    [E]+[C]=[E]_0+[C]_0, ~[C]+[P]=[C]_0+[P]_0, ~ 
    \kappa_2^+  [C]       = \kappa_2^-  [P][E], ~
    [S]=[S]^e, 
\end{equation*}
whose solution is also different from the original equilibrium state. 
Moreover, since $d[S]/dt=0$, the Conservation law 2, $[S]+[C]+[P]=Const$, for the full MM reactions holds too. So that no modification has to be made to our previous derivations on PEA-MM2 without conservation laws. The only attention needs be paid to the initial values to avoid trivial solutions. In contrast, for PEA-MM1, the inclusion of $[S]+[C]+[P]=Const$ brings an extra constraint to the reduced system and will cause a contradiction.

On the other hand, once the original conservation laws are utilized in the derivation of PEA-reduced models for MM reactions, the corresponding thermodynamics of reduced models become intricate that beyond the scope of the current study. Here we emphasize that the adoption of Conservation law 1 leads to the usual PEA-reduced dynamics of MM reactions. 
\begin{rem}
Substituting the PEA relation \eqref{R1pm} into the Conservation law 1, $[E]+[C]=Const$, yields instantaneously the expression of the complex as $[C] = \frac{([E]_0+[C]_0) [S]}{K_e  + [S]}$ and then the generation rate of the product as 
\begin{equation}
\label{gen rate}
\frac{{d[P]}}{{dt}} = (\kappa _2 ^ +   + \kappa _2 ^ -  [P])\frac{{([E]_0+[C]_0) [S]}}{{K_e  + [S]}} - \kappa _2 ^ -  ([E]_0+[C]_0) [P],
\end{equation}
where $K_e\equiv \kappa_1^-/\kappa_1^+$ is known as the Michaelis-Menten saturation coefficient. For irreversible MM reactions with $\kappa_2^-=0$, the generation rate in Eq. \eqref{gen rate} becomes $\frac{{d[P]}}{{dt}} = \kappa_2^+ \frac{([E]_0+[C]_0) [S]}{K_e  + [S]}$, which recovers the classic results obtained by Michaelis and Menten. 
\end{rem}

Interestingly, the break down of old conservation laws and emergence of new ones reveal a rich algebraic structure of the reduced models for such simple MM reactions. Fruitful new results are naturally expected for more complex CRNs. For example, as to open CRNs, Polettini and Esposito \cite{polettini2014irreversible} reported that the sum of numbers of independent affinities cycles and broken conservation laws equal to the number of chemostats. Please see Refs. \cite{Rao2016Nonequilibrium, polettini2014irreversible} for their connections to the thermodynamic properties. 

\subsection{Thermodynamics for QSSA-reduced MM reactions}
\label{QSSA_thermodynamics_MM}

Different from PEA, QSSA assumes that the synthesis and decomposition rates of the complex $C$ equal, or equivalently, the third expression of the ODEs in \eqref{mass-action-MM} is reduced to an algebraic equation, 
\begin{equation}
\label{QSSA}
(\kappa _1 ^ +  [S] + \kappa _2 ^ -  [P])[E] = (\kappa _1 ^ -   + \kappa _2 ^ +  )[C].
\end{equation} 
In this stage, we have four reduced ODEs 
\begin{subequations}
\label{mass-action-MM4}
\begin{align}
 \frac{{d[S]}}{{dt}}&=  - \kappa _1 ^ +  [S][E] + \kappa _1 ^ -  [C], \label{mass-action-MM41}\\ 
 \frac{{d[E]}}{{dt}}&=  0, \label{mass-action-MM42}\\ 
 \frac{{d[C]}}{{dt}}&= 0, \label{mass-action-MM43}\\ 
 \frac{{d[P]}}{{dt}}&= \kappa_2^+  [C] - \kappa_2^-  [P][E], \label{mass-action-MM44}
 \end{align}
\end{subequations}
with the initial concentrations $([S]_0, [E]_0, [C]_0, [P]_0)^T$ and the QSSA relation in Eq. \eqref{QSSA}. 
Now we proceed to discuss the reduced MM reactions in QSSA-MM1 and QSSA-MM2 separately. 

\subsubsection{Thermodynamics for QSSA-MM1: Use both algebraic relation and conservation law}

Notice by applying QSSA-1 to MM reactions, all old conservation laws have been broken. Instead, a new conservation law emerges, $[S]+[P]=Const$. Just by coincidence, in this case, we have only one algebraic relation \eqref{QSSA} and two differential equations \eqref{mass-action-MM41} and \eqref{mass-action-MM44} concerning four unknown variables. In order to make the reduced model be uniquely solvable, the Conservation law 1, $[E]+[C]=Const$, is introduced as a complementary constraint. 

By neglecting the ODEs in \eqref{mass-action-MM42}-\eqref{mass-action-MM43} and substituting Eq. \eqref{QSSA} into the Conservation law 1, we obtain the expressions of the complex $C$ and enzyme $E$ represented via the substrate $S$ and product $P$, 
\begin{equation}
\label{CE}
 [C] = ([E]_0+[C]_0) \frac{{\kappa _1 ^ +  [S] + \kappa _2 ^ -  [P]}}{{\kappa _1 ^ -   + \kappa _2 ^ +   + \kappa _1 ^ +  [S] + \kappa _2 ^ -  [P]}}, \quad
 [E] = \frac{{(\kappa _1 ^ -   + \kappa _2 ^ +  )[C]}}{{\kappa _1 ^ +  [S] + \kappa _2 ^ -  [P]}} = ([E]_0+[C]_0) \frac{{\kappa _1 ^ -   + \kappa _2 ^ +  }}{{\kappa _1 ^ -   + \kappa _2 ^ +   + \kappa _1 ^ +  [S] + \kappa _2 ^ -  [P]}}.
\end{equation}
Upon substituting the algebraic equations of $[C]$ and $[E]$ above into the remaining ODEs in \eqref{mass-action-MM41} and \eqref{mass-action-MM44}, we obtain the QSSA-reduced dynamics of MM reactions in terms of $[S]$ and $[P]$ as 
\begin{subequations}
\label{QSSA-MM1}
\begin{align}
 \frac{{d[S]}}{{dt}}&= ([E]_0+[C]_0) \frac{{\kappa _1 ^ -  \kappa _2 ^ -  [P] - \kappa _1 ^ +  \kappa _2 ^ +  [S]}}{{\kappa _1 ^ -   + \kappa _2 ^ +   + \kappa _1 ^ +  [S] + \kappa _2 ^ -  [P]}}, \\ 
 \frac{{d[P]}}{{dt}}&= ([E]_0+[C]_0) \frac{{\kappa _1 ^ +  \kappa _2 ^ +  [S] - \kappa _1 ^ -  \kappa _2 ^ -  [P]}}{{\kappa _1 ^ -   + \kappa _2 ^ +   + \kappa _1 ^ +  [S] + \kappa _2 ^ -  [P]}},
 \label{QSSA-MM1-2}
 \end{align}
\end{subequations}
where the initial condition is $([S],[P])^T = ([S]_0, [P]_0)^T$. 
The Eqs. \eqref{CE}-\eqref{QSSA-MM1} together are called the QSSA-MM1 model, whose steady state is governed by 
\begin{equation*}
    [E]+[C]=[E]_0+[C]_0, ~[S]+[P]=[S]_0+[P]_0, ~ 
    (\kappa _1 ^ +  [S] + \kappa _2 ^ -  [P])[E] = (\kappa _1 ^ -   + \kappa _2 ^ +  )[C], ~
    \kappa _1 ^ +  \kappa _2 ^ +  [S] = \kappa _1 ^ -  \kappa _2 ^ -  [P].
\end{equation*}
This steady state is usually different from that of the full model.

\begin{rem}
We remark that the generation rate of the product $P$ can be recast into  
\begin{equation}
\frac{d[P]}{dt} = ([E]_0+[C]_0) \frac{\kappa _2 ^ +  \frac{{[S]}}{{K_s }} - \kappa _1 ^ -  \frac{{[P]}}{{K_p }}}{1 + \frac{{[S]}}{{K_s }} + \frac{{[P]}}{{K_p }}},
\end{equation}
where the constants $K_s=(\kappa _1 ^ -   + \kappa _2 ^ +  )/{{\kappa_1^+  }}$ and $K_p=(\kappa _1 ^ -   + \kappa _2 ^ + )/{\kappa_2^-}$. 
Analogous to the case discussed in PEA, when considering the irreversible MM reactions with $\kappa_2^-=0$, this generation rate recovers to the classic conversion rate of $S$ into $P$, $\frac{d[P]}{dt} = \kappa _2 ^ +  \frac{{([E]_0+[C]_0) [S]}}{{K_s  + [S]}}$. 
Note that the generation rate of the product of QSSA-reduced equation is slightly different from that of PEA: the denominator of the generation rate obtained by QSSA is $K_s  = (\kappa _1 ^ -   + \kappa _2 ^ +  )/\kappa _1 ^ +  $, while that of PEA is $K_e  = \kappa _1 ^ -  /\kappa _1 ^ +$. 
\end{rem}

We proceed to derive the entropy change rate for the QSSA-MM1 model as follows: 
\begin{align*}
\frac{d}{dt}\widetilde{Ent}(t)
&=-\ln\left(\frac{[E]}{[C]}\right) \frac{d[E]}{dt} -\ln\left(\frac{[S]}{[P]}\right)\frac{d[S]}{dt}\\
&=\ln\left( \frac{ \kappa _1 ^ -   + \kappa _2 ^ + }{ \kappa _1 ^ +  [S] + \kappa _2 ^ -  [P] } \right) \frac{(\kappa _1 ^ -   + \kappa _2 ^ +)(\kappa _1 ^ + - \kappa _2 ^ -) ([E]_0+[C]_0)}{(\kappa _1 ^ -   + \kappa _2 ^ +   + \kappa _1 ^ +  [S] + \kappa _2 ^ -  [P])^2} \frac{d[S]}{dt}
-\ln\left(\frac{[S]}{[P]}\right)\frac{d[S]}{dt}\\
&=([E]_0+[C]_0) \frac{\kappa _1 ^ +  \kappa _2 ^ +  [S] - \kappa _1 ^ -  \kappa _2 ^ -  [P] }{\kappa _1 ^ -   + \kappa _2 ^ +   + \kappa _1 ^ +  [S] + \kappa _2 ^ -  [P]}
\ln \left[ \left(\frac{[S]}{[P]} \right) 
\left( \frac{ \kappa _1 ^ +  [S] + \kappa _2 ^ -  [P] }{ \kappa _1 ^ -   + \kappa _2 ^ + } \right)^{\left( \frac{(\kappa _1 ^ -   + \kappa _2 ^ +)(\kappa _1 ^ + - \kappa _2 ^ -) ([E]_0+[C]_0)}{(\kappa _1 ^ -   + \kappa _2 ^ +   + \kappa _1 ^ +  [S] + \kappa _2 ^ -  [P])^2} \right) } 
\right],
\end{align*}
from which the entropy production rate and entropy flow rate are separated and obtained as 
\begin{align}
\widetilde{epr}(t)&=([E]_0+[C]_0) \frac{{\kappa _1 ^ +  \kappa _2 ^ +  [S] - \kappa _1 ^ -  \kappa _2 ^ -  [P]}}{{\kappa _1 ^ -   + \kappa _2 ^ +   + \kappa _1 ^ +  [S] + \kappa _2 ^ -  [P]}}
\ln \frac{\kappa _1 ^ +  \kappa _2 ^ + [S]}{\kappa _1 ^ - \kappa _2 ^ -  [P]} \geq 0, \\
\widetilde{J^f}(t)&=([E]_0+[C]_0) \frac{\kappa _1 ^ +  \kappa _2 ^ +  [S] - \kappa _1 ^ -  \kappa _2 ^ -  [P] }{\kappa _1 ^ -   + \kappa _2 ^ +   + \kappa _1 ^ +  [S] + \kappa _2 ^ -  [P]}
\ln \left[ \left(\frac{ \kappa_1^- \kappa_2^- }{\kappa _1 ^ +\kappa_2^+ } \right) 
\left( \frac{ \kappa _1 ^ +  [S] + \kappa _2 ^ -  [P] }{ \kappa _1 ^ -   + \kappa _2 ^ + } \right)^{\left( \frac{(\kappa _1 ^ -   + \kappa _2 ^ +)(\kappa _1 ^ + - \kappa _2 ^ -) ([E]_0+[C]_0)}{(\kappa _1 ^ -   + \kappa _2 ^ +   + \kappa _1 ^ +  [S] + \kappa _2 ^ -  [P])^2} \right) } 
\right].
\end{align}
Analogously, the free energy dissipation rate is calculated as 
\begin{align}
\label{fd-QSSAMM1}
&\widetilde{f_d}(t) 
=-\ln\left( \frac{[E][C]^e}{[E]^e[C]} \right) \frac{d[E]}{dt}
-\ln\left(\frac{[S][P]^e}{[S]^e[P]}\right)\frac{d[S]}{dt}\nonumber\\
&= ([E]_0+[C]_0) \frac{\kappa _1 ^ +  \kappa _2 ^ +  [S] - \kappa _1 ^ -  \kappa _2 ^ -  [P] }{\kappa _1 ^ -   + \kappa _2 ^ +   + \kappa _1 ^ +  [S] + \kappa _2 ^ -  [P]}
\ln \left[ \left( \frac{[S][P]^e}{[S]^e[P]} \right) 
\left( \frac{ \kappa_1^- (\kappa _1 ^ +  [S] + \kappa _2 ^ -  [P]) }{ (\kappa _1 ^ -   + \kappa _2 ^ +) \kappa_1^+[S]^e } \right)^{\left( \frac{(\kappa _1 ^ -   + \kappa _2 ^ +)(\kappa _1 ^ + - \kappa _2 ^ -) ([E]_0+[C]_0)}{(\kappa _1 ^ -   + \kappa _2 ^ +   + \kappa _1 ^ +  [S] + \kappa _2 ^ -  [P])^2} \right) } 
\right],
\end{align}
recalling that the tilde $\widetilde{\mathcal{C}}$ is utilized to denote the QSSA-reduced results. Based on the above formulas, we can see that the entropy production rate $\widetilde {epr}(t)$ of QSSA-MM1 model is non-negative, while the free energy dissipation rate $\widetilde{f_d}(t)$ has an indefinite sign.

\subsubsection{Thermodynamics for QSSA-MM2: Use no algebraic relation}

Via neglecting the algebraic equation $(\kappa _1 ^ +  [S] + \kappa _2 ^ -  [P])[E] = (\kappa _1 ^ -   + \kappa _2 ^ +  )[C]$, and adopting the corresponding differential equations $d[E]/dt=-d[C]/dt=0$ with initial values $([E]^e, [C]^e)^T$, we conclude that in addition to the original two conservation laws $[E]+[C]=Const$ and $[S] + [C] + [P] = Const$, two new conservation laws emerge for the QSSA-MM2, i.e. $E(t)=[E]^e$ and $[C](t)=[C]^e$. As a consequence, the governing equations of the QSSA-MM2 read 
\begin{subequations}
\label{QSSA-MM2}
\begin{align}
 \frac{{d[S]}}{{dt}}&= -\kappa_1^+  [S][E]^e + \kappa_1^-  [C]^e, \\ 
 \frac{{d[P]}}{{dt}}&= \kappa_2^+  [C]^e - \kappa_2^-  [P][E]^e,
 \end{align}
\end{subequations}
where the initial values are $([S]_0,[P]_0)^T$. It is direct to find that the steady state of QSSA-MM2 satisfies that 
\begin{equation*}
    [S] = \frac{\kappa_1^-  [C]^e}{\kappa_1^+[E]^e}, ~
    [P]=\frac{\kappa_2^+  [C]^e}{\kappa_2^-  [E]^e}, ~ 
    [E]\equiv [E]^e, ~
    [C]\equiv [C]^e, 
\end{equation*}
whose solution is the same as the original equilibrium state of the full model. 

\begin{rem}
Due to the essential distinction between closed and open CRNs in the aspect of exchanging matters with environment, their dynamics and thermodynamics are usually studied separately. Surprisingly, the QSSA-MM2 model in Eq. \eqref{QSSA-MM2} corresponds to an open CRNs as 
$S \xrightleftharpoons[{\kappa_1^- [C]^e}]{{\kappa_1^+ [E]^e}} \phi \xrightleftharpoons[{\kappa_2^- [E]^e}]{{\kappa_2^+ [C]^e}} P$, with $\phi$ denoting a zero species, and the content above/below the arrows denoting the rates of four pseudo-reactions. This example also leads naturally to our upcoming work on the reduced thermodynamics of open CRNs. The relationship between open CRNs and the pseudo-reactions within a Continuous-Flow Stirred-Tank Reactor (CFSTR), including the model network, kinetics and induced ODEs, is illustrated comprehensively in Ref.  \cite{feinberg2019foundations}.
\end{rem}

To further analyze the thermodynamics of this QSSA-MM2 model, we apply the thermodynamics of general reactions reduced via QSSA-2 in Sect. \ref{QSSA thermodynamics} to MM reactions and obtain
\begin{subequations}
\label{MM_thermo4}
\begin{align}
\widetilde{J^f}(t)&=(\kappa_1^+  [S][E]^e - \kappa_1^-  [C]^e)\ln \frac{\kappa_1^-  [C]^e}{\kappa_1^+[E]^e } + 
(\kappa_2^+  [C]^e - \kappa_2^-  [P][E]^e)\ln \frac{\kappa_2^-  [E]^e}{\kappa_2^+  [C]^e}, \\
\widetilde{epr}(t)&=
(\kappa_1^+  [S][E]^e - \kappa_1^-  [C]^e)\ln \frac{\kappa_1^+  [S][E]^e }{\kappa_1^-  [C]^e} + 
(\kappa_2^+  [C]^e - \kappa_2^-  [P][E]^e)\ln \frac{\kappa_2^+  [C]^e}{\kappa_2^-  [P][E]^e} \ge 0 \\
\widetilde{f_d}(t)&=(\kappa_1^+  [S][E]^e - \kappa_1^-  [C]^e)\ln \frac{[S]}{[S]^e} + 
(\kappa_2^+  [C]^e - \kappa_2^-  [P][E]^e)\ln \frac{[P]^e}{[P]}=\widetilde{epr}(t) \ge 0.
\end{align}
\end{subequations}
The above expressions show that the entropy production rate of the QSSA-MM2 model is preserved to be non-negative, $\widetilde{epr}(t) \geq 0$; and $\widetilde{f_d}(t) =\widetilde{epr}(t)$. 

\subsection{Numerical illustration}
\begin{figure}[h]
    \centering
    \includegraphics[width=0.7\linewidth, height=1.0\linewidth]{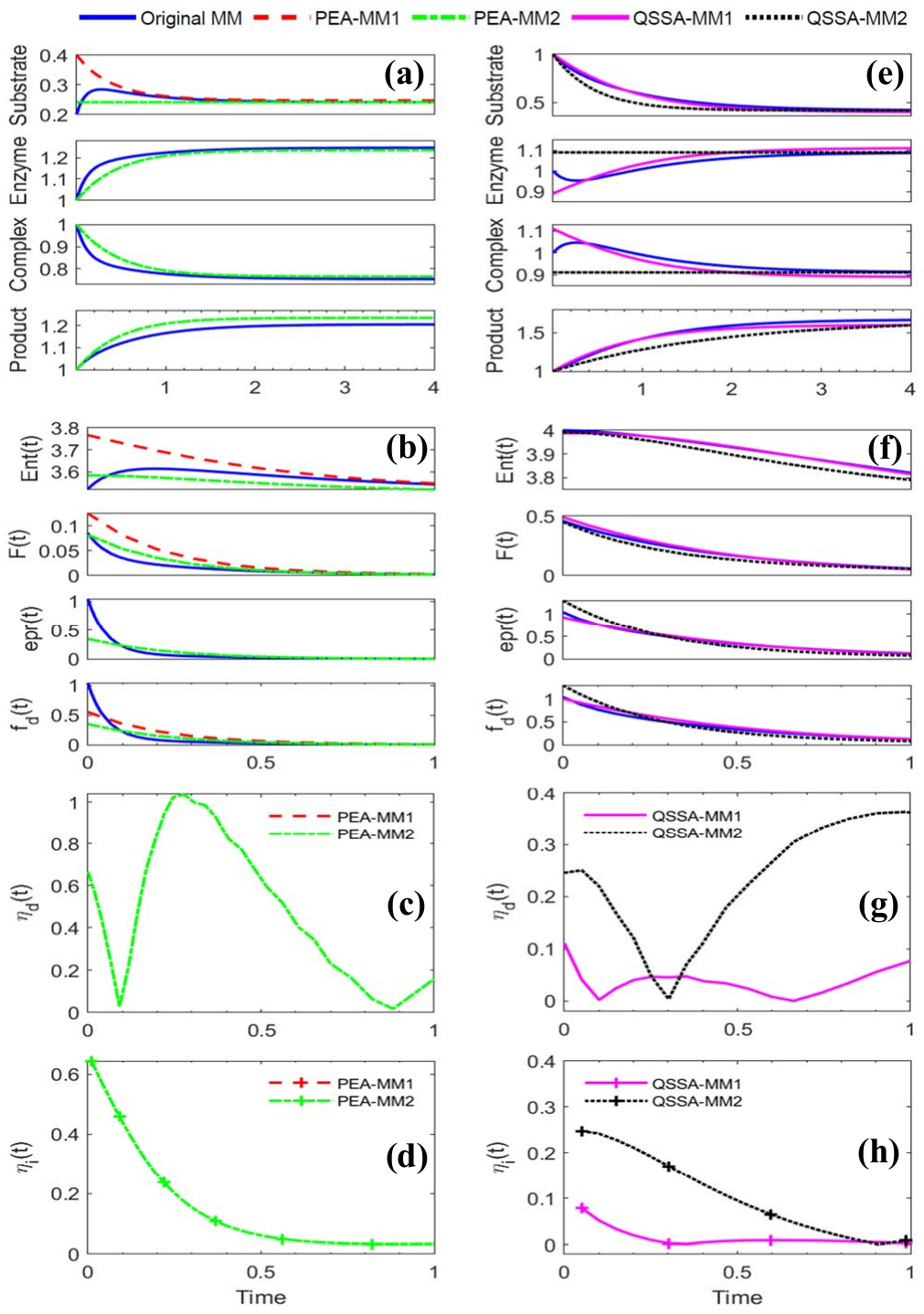}
    \caption{Illustration on the dynamics and thermodynamics of the full model (blue solid lines) and reduced models by either PEA-MM1 (red dashed lines) and PEA-MM2 (green dash-dot lines), or QSSA-MM1 (magenta solid lines) and QSSA-MM2 (black dotted lines) for MM reactions. From the left four panels (a-d) to the right panels (e-h), the results for PEA and QSSA models are shown respectively. 
    (a, e) From the top down, trajectories of the substrate $[S](t)$, enzyme $[E](t)$, complex $[C](t)$ and product $[P](t)$ for the full model and reduced models are compared and plotted in the four subplots. So are the entropy $Ent(t)$, free energy $F(t)$, entropy production rate $epr(t)$ and free energy dissipation rate $f_d(t)$ in (b, f). The accuracy of four model-reduction methods is illustrated through the relative difference of instantaneous entropy production rate (c, g) $\eta_d(t)$  and of total entropy production during time interval $[0, t]$ (d, h) $\eta_i(t)$ .
    For all plots, the initial values and rate constants are taken as (a-d) $([S]_0, [E]_0, [C]_0, [P]_0) = (0.2, 1, 1, 1)$, $(\kappa_1^+, \kappa_1^-, \kappa_2^+, \kappa_2^-)=(5, 2, 1, 0.5)$; and (e-h) $([S]_0, [E]_0, [C]_0, [P]_0) = (1, 1, 1, 1)$, $(\kappa_1^+, \kappa_1^-, \kappa_2^+, \kappa_2^-)=(2,1,1,0.5)$ correspondingly.
    }
    \label{Fig1}
\end{figure}

To provide a better understanding on the thermodynamic results we obtained for PEA and QSSA methods, we perform numerical calculations of the MM reactions as an illustration. As shown by the trajectories of the substrate $[S](t)$, enzyme $[E](t)$, complex $[C](t)$ and product $[P](t)$ in Fig. \ref{Fig1}(a,e), the reduced models by either PEA or QSSA methods offer quite good approximations of original solutions to the full model, as long as the fast equilibration assumption on reactions/reactants holds. Overall, the performance of PEA-1 (or QSSA-1) looks better than that of PEA-2 (or QSSA-2) from the dynamic aspect, which could be attributed to the fact using algebraic relations for reduced variables.

\begin{figure}[h]
    \centering
    \includegraphics[width=0.8\linewidth]{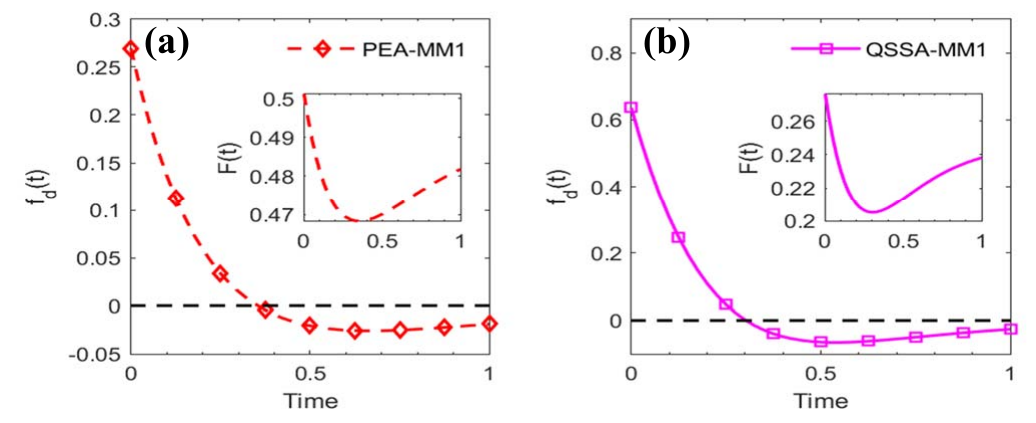}
    \caption{Illustration on the negative part of the free energy dissipation rate $f_d(t)$ for reduced models of MM reactions by either (a) PEA-1 or (b) QSSA-1. The trajectory of free energy dissipation rate is marked by the red dotted line with diamonds for PEA-MM1 and by the magenta solid line with squares for QSSA-MM1.  The corresponding free energy function is shown in the \textit{Inset}.
    The initial values and rate constants are (a) $([S]_0, [E]_0, [C]_0, [P]_0) = (2, 1, 1, 1)$, $(\kappa_1^+, \kappa_1^-, \kappa_2^+, \kappa_2^-)=(2,1,1,0.5)$; and (b) $([S]_0, [E]_0, [C]_0, [P]_0) = (0.2, 1, 1, 0.2)$, $(\kappa_1^+, \kappa_1^-, \kappa_2^+, \kappa_2^-)=(5,2,1,0.5)$ respectively. Remark that, the above two sets of parameters are the same as those in Fig. \ref{Fig1}, except for  initial values of the substrate or product.}
    \label{Fig2}
\end{figure}

As to the thermodynamic behaviors of the reduced models which we are more concerned about in the current study, we make an exploration based on four thermodynamic quantities -- entropy $Ent(t)$, free energy $F(t)$, entropy production rate $epr(t)$ and free energy dissipation rate $f_d(t)$. The difference between the latter two quantities for both PEA-1 and QSSA-1 models can be directly seen from Fig. \ref{Fig1}(b,f). Furthermore, the negative part of the free energy dissipation rate for both PEA-1 and QSSA-1 models is illustrated in Fig. \ref{Fig2}, which highlights the value of our study by exploring the reduced models from a thermodynamic aspect.

Since in all reduced models, the entropy production rate keeps non-negative, we introduce the \emph{relative difference of instantaneous entropy production rate} between the original model $epr(t)$ and the reduced model $\widehat{epr}(t)$ as 
\begin{equation}
    \eta_d(t)=\frac{|epr(t)-\widehat{epr}(t)|}{epr(t)},
\end{equation}
to evaluate the accuracy of various methods of model reduction from a thermodynamic aspect. As illustrated in Fig. \ref{Fig1}(c,g), $\eta_d(t)$ exhibits a non-monotonic dependence on the time. The smaller this value is, the better the approximate method is in aspect of thermodynamics. Leave alone the coincident agreement of $\eta_d(t)$ for PEA-1 and for PEA-2 in the case of MM reactions, the relative difference of instantaneous entropy production rate $\eta_d(t)$ for QSSA-1 is much smaller than that of QSSA-2 during almost the whole time region under study. This observation agrees with previous dynamic results. In addition, by introduce the \emph{relative difference of total entropy production during a time interval $[t_1, t_2]$}
\begin{equation}
    \eta_i(t_2;t_1)=\frac{|\int_{t_1}^{t_2} epr(t) dt - \int_{t_1}^{t_2}  \widehat{epr}(t) dt|}{\int_{t_1}^{t_2} epr(t) dt},
\end{equation}
the comparison can be done in a more smooth way. A more useful case with fixed starting time point $t_1=0$, and arbitrary ending time $t_2 \geq 0$ becomes $\eta_i{(t)}={|\int_{0}^{t} epr(\tau) d\tau - \int_{0}^{t}  \widehat{epr}(\tau) d\tau|}/{\int_{0}^{t} epr(\tau) d\tau}$, which is drawn in Fig. \ref{Fig2}(d,h) for MM reactions. 


\section{Conclusion and Discussion}
Previous studies have mainly focused on the accuracy of solutions before and after applying model reduction techniques, but this paper took a thermodynamic perspective. 
We have established a quantitative connection between the thermodynamic quantities, including the entropy production rate, free energy dissipation rate, and entropy flow rate, in the original reversible chemical mass-action equations and the reduced models obtained through either PEA or QSSA, via Eqs. \eqref{PEA1:Jf}-\eqref{QSSA2:fd} and Table \ref{table1}. 
The results revealed that both PEA and QSSA methods do not necessarily preserve the thermodynamic structure of the original full model during the reduction process. Specifically, the loss of non-negativity of the free energy dissipation rate was observed when algebraic relations were used instead of differential equations. 
The above general results were carefully validated though the application to MM reactions as a concrete example, both theoretically and numerically. 
Even such a simple model has clearly demonstrated the complicated thermodynamic structures of its reduced mechanisms, which provides a prototype for further discussions on thermodynamics for the reduction of more complex CRNs. 
Overall, this study contributes to a better understanding on the limitations and potential drawbacks of PEA and QSSA in preserving the thermodynamic properties of chemical reactions. It emphasizes the importance of considering thermodynamic principles when applying model reduction techniques and suggests the necessity for further research in this direction.

With regard to potential theoretical extensions to the current study, we first mention the maximum entropy principle (MEP), which can deduce the most likely concentrations (or probability distribution in mesoscopic scale) as the one that maximizes the entropy function with respect to some given constraints. These optimal concentrations can be used to derive several more macroscopic variables, such as the moments, and therefore reduce the number of unknown variables \cite{smadbeck2013closure,hong2013simple}. 
The next extension concerns about irreversible chemical reactions. Recall that in the previous sections we treat all reactions to be reversible. However, employing irreversible reactions usually yields simpler dynamics and thus is preferably adopted. From the perspective of model reduction, a reaction can be viewed as irreversible when the backward reaction rate is significantly smaller than the forward one \cite{gorban2013thermodynamics}. Moreover, the thermodynamics of reduced models for open CRNs shall be studied in parallel with the closed CRNs in future works. 

Taking a broader view on model reduction or coarse-graining of CRNs, the current study is closely relevant to the multi-scale descriptions of CRNs such as chemical master equation, chemical Langevin equation, chemical mass-action equation, etc. There are many different ways to simplify these models \cite{pigolotti2008coarse,puglisi2010entropy,esposito2012stochastic,kawaguchi2013fluctuation,bo2017multiple,busiello2019entropy,busiello2019entropynpj,yu2021inverse,yu2022state,DeGiuli_2022}, and the thermodynamics of the reduced models are expected to present much richer content, among which the mean-field theory \cite{DeGiuli_2022} and the renormalization group \cite{pigolotti2008coarse,yu2021inverse,yu2022state} deserve special attentions. 

From the perspective of macroscopic non-equilibrium thermodynamics, for instance the CDF (conservation-dissipation formalism of irreversible thermodynamics \cite{peng2021recent}) and GENERIC (general equation for the non-equilibrium reversible-irreversible coupling \cite{pavelka2018multiscale}), the model reduction corresponds to the inverse problem of model construction based on some known principles. Taking CDF as an example, it is rooted on the conservation laws like mass, momentum and energy conservation, and then the choice of conserved and dissipative variables, which correspond to the slow and fast variables of a physical system respectively. CDF builds new extended models that are thermodynamically compatible and mathematically stable by introducing suitable fast variables, while model reduction eliminates the fast variables and keeps the slow ones of the system as leading terms. As a consequence, it is natural to require the reduced model still possess nice thermodynamic properties as the original full model.


\section*{Declaration of Competing Interest}
The authors declared no competing interests.

\section*{Acknowledgements}
The authors acknowledged the financial supports from the National Natural Science Foundation of China (Grants No. 12205135, 21877070), the Natural Science Foundation of Fujian Province of China (2020J05172), Guangdong Basic and Applied Basic Research Foundation (2023A1515010157), and the Startup Funding of Minjiang University (mjy19033).

\bibliographystyle{unsrt}%
\bibliography{fp}
\end{document}